 \documentclass[10pt,pre,twocolumn,longbibliography,amsmath,amssymb]{revtex4-1}
 
\usepackage{stmaryrd} 
\usepackage{hyperref} 

 \usepackage{bm}
 \usepackage{graphicx}
 \usepackage{latexsym}
 \usepackage{enumerate}

\usepackage{upgreek}

\usepackage[mathscr]{euscript} 

\newcommand{\muSI}{\upmu_0}

\newcommand{\iotabar}{\mbox{$\,\iota\!\!$-}}

\newcommand{\vrm}[1]{\bm{\mathrm{#1}}} 
\newcommand{\jump}[1]{\left\llbracket  #1 \right\rrbracket} 
\newcommand{\lbrak}{\left\langle}
\newcommand{\rbrak}{\right\rangle}
\renewcommand{\d}{d} 

\newcommand{\grad}{ \mbox{\boldmath\(\nabla\)}}
\newcommand{\divv}{ \mbox{\boldmath\(\nabla\cdot\)}}
\newcommand{\curl}{ \mbox{\boldmath\(\nabla\times\)}}

\newcommand{\esub}[1]{{\bf e}_{#1}}

\newcommand{\dotv}{  \mbox{\boldmath\(\cdot\)} }

\newcommand{\cross}{  \mbox{\boldmath\(\times\)} }

\newcommand{\half}{\mbox{\small\(\frac{1}{2}\)} }
\newcommand{\const}{{\mathrm{const}}}
\renewcommand{\eqref}[1]{Eq.~(\ref{#1})}

\newcommand{\etal}{\emph{et al.}\ }

\newcommand{\gammas}{\gamma_{\rm S}}
\newcommand{\xbdy}{x_{\rm bdy}}
\newcommand{\abdy}{a}

\newcommand{\Aalpha}{\mathscr{A}}

\newcommand{\psiU}{\psi_{\rm U}}
\newcommand{\psiav}{\overline{\psi}}
\newcommand{\psibdy}{\psi_a}
\newcommand{\psicut}{\psi_{\rm cut}}
\newcommand{\psitld}{\widetilde\psi}
\newcommand{\psihat}{\widehat\psi}
\newcommand{\Fav}{\overline{F}}

\newcommand{\AH}{\vrm{A}^{\rm H}}

\newcommand{\BH}{\vrm{B}_{\rm H}}

\newcommand{\WSigma}{\mathcal{W}_{\Sigma}}

\newcommand{\cosav}{{\lbrak\cos\mu_0 x\rbrak_0}}

\begin{document}

\title{Multi-region relaxed magnetohydrodynamics in plasmas with slowly changing boundaries --- resonant response of a plasma slab} 

\author{R.~L. Dewar}
 \email{robert.dewar@anu.edu.au}
\affiliation{Centre for Plasmas and Fluids, Research School of Physics \& Engineering, The Australian National University, Canberra, ACT 2601, Australia}

\author{S.~R. Hudson}
 \email{shudson@pppl.gov}
 \affiliation{Princeton Plasma Physics Laboratory, PO Box 451, Princeton NJ 08543, USA}

\author{A. Bhattacharjee}
 \email{abhattac@pppl.gov}
 \affiliation{Princeton Plasma Physics Laboratory, PO Box 451, Princeton NJ 08543, USA}
 
 \author{Z. Yoshida}
 \email{yoshida@ppl.k.u-tokyo.ac.jp}
 \affiliation{Graduate School of Frontier Sciences, University of Tokyo, Kashiwa, Chiba 277-8561, Japan}

\date{\today}

\begin{abstract}
The adiabatic limit of a recently proposed dynamical extension of Taylor relaxation, \emph{multi-region relaxed magnetohydrodynamics} (MRxMHD) is summarized, with special attention to the appropriate definition of relative magnetic helicity. The formalism is illustrated using a simple two-region, sheared-magnetic-field model similar to the Hahm--Kulsrud--Taylor (HKT) rippled-boundary slab model. In MRxMHD a linear Grad--Shafranov equation applies, even at finite ripple amplitude. The adiabatic switching on of boundary ripple excites a shielding current sheet opposing reconnection at a resonant surface. The perturbed magnetic field as a function of ripple amplitude is calculated by invoking conservation of magnetic helicity in the two regions separated by the current sheet. At low ripple amplitude ``half islands'' appear on each side of the current sheet, locking the rotational transform at the resonant value. Beyond a critical amplitude these islands disappear and the rotational transform develops a discontinuity across the current sheet.
\end{abstract}

\maketitle

\section{Introduction}\label{sec:intro}

The deficiencies of ideal magnetohydrodynamics (MHD) for describing typical fusion plasmas arise from its assumption of infinite electrical conductivity, which implies ``frozen-in'' magnetic flux \cite{Newcomb_58}, and also its assumption of zero thermal conductivity, which implies frozen-in entropy (i.e. a thermodynamically adiabatic equation of state applying in each fluid element). The problem with frozen-in entropy is obvious---thermal conductivity along magnetic field lines is in fact extremely high. The problem with frozen-in flux is that it precludes changes in magnetic-field-line topology through such reconnection phenomena as the growth of magnetic islands at resonant magnetic surfaces.
Such islands may be excited by breaking axisymmetry using external coils, as in stellarators or tokamaks with applied resonant magnetic perturbations (RMPs \cite{Evans_etal_06}), 
or through spontaneous tearing mode instability \cite{White_etal_77}. 

To allow for magnetic reconnection and parallel thermal equilibration, while retaining the non-dissipative character of ideal MHD, we use a \emph{multiregion relaxation} (MRx) model where complete Taylor relaxation \cite{Taylor_74} occurs only within subregions of the plasma. These  \emph{relaxation regions} are separated by a number (in principle, many) \emph{interfaces} \cite{Hole_Hudson_Dewar_06}, or transport barriers, that act like thin layers of ideal plasma where the ideal-MHD invariants are preserved, unlike the relaxation regions where the only magnetic invariants are helicity and total fluxes. These interfaces frustrate the total relaxation postulated in the original Taylor model,
which cannot model the peaked current and pressure profiles sought in fusion plasma physics. 

\begin{figure}[htbp]
\begin{center}
	\includegraphics[width=0.45\textwidth]{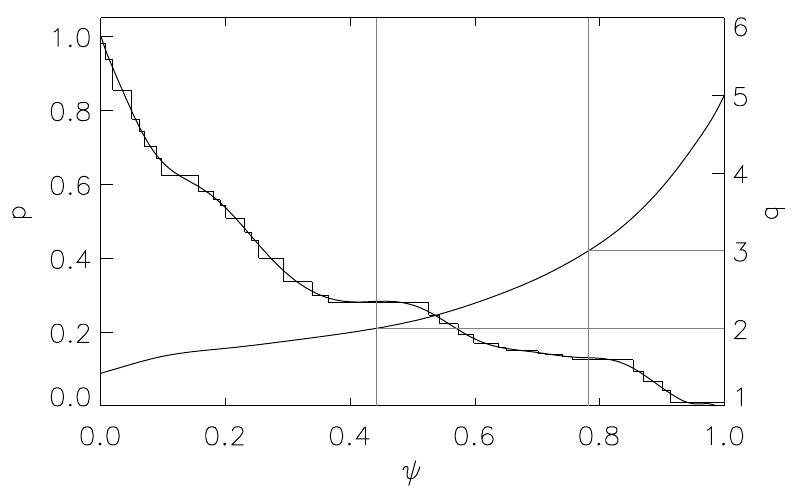}
\caption{Comparison of a smooth pressure ($p$) profile from a DIII-D reconstruction, using the STELLOPT code, and the stepped $p$-profile used in a SPEC calculation of the corresponding 3-D equilibrium. Also, plotted is the inverse rotational transform  $\equiv$ safety factor $q$. [Reproduced with permission from Phys. Plasmas \textbf{19}, 112502 (2012).]}
\label{fig:DIIID}
\end{center}
\end{figure}

The MRx model is much more flexible, and has 
been used as the theoretical basis for the three-dimensional (3-D) equilibrium code SPEC \cite{Hudson_etal_12b}. This has
already had success \cite[Sec.~IV. E ]{Hudson_etal_12b} in modeling MHD equilibria using experimental data from DIII-D tokamak shots, where RMP coils were used to stochasticize the outer region of the plasma in order to ameliorate edge-localized modes (ELMs) \cite{Evans_etal_06}. Figure~\ref{fig:DIIID} reproduces Fig.~7 of \cite{Hudson_etal_12b}, showing the smooth pressure profile produced by a STELLOPT \cite{Lazerson_etal_12}/VMEC \cite{Hirshman_Betancourt_91} reconstruction to fit the DIII-D data. Also shown is a closely approximating stepped-pressure profile used in a SPEC calculation with multiple relaxation regions and the same rippled boundary as used by VMEC. Evident in both are wide regions around the $q = 2$ and $q = 3$ surfaces where  the pressure profile is flattened, presumably due to the presence of field-line chaos and islands generated by the RMP coils, as verified in the SPEC-produced Fig.~8 of \cite{Hudson_etal_12b}. However, while VMEC can produce adequate macroscopic fits at a specific time in the discharge, it is based on ideal MHD so it cannot resolve field-line chaos and islands, and hence cannot model the \emph{development} of equilibria exhibiting field-line chaos. The present paper presents the theoretical basis for believing that SPEC can potentially do this.  

Taylor \cite{Taylor_86} postulates \emph{macroscopic} relaxation to a force-free magnetic field $\vrm{B}$ as due to turbulent fluctuations of short (\emph{microscale}) wavelength, scaling as the square root of the resistivity. Other fluctuation arguments for ubiquity of relaxation may be advanced \cite{Qin_Liu_Li_Squire_12}.  

However, the viewpoint we adopt in this paper is 
that field-line chaos (``stochasticity'') leads to relaxation by entangling \cite{Rusbridge_91} the microscopic flux tubes that each carry their own conserved magnetic helicity, combined with a reconnection mechanism that leaves only the total magnetic helicity conserved, as assumed by Taylor. A 
related, dynamical-systems-based, line of reasoning advanced by Hudson \etal \cite{Hudson_etal_12b} in justification of our MRx model is that a partition of the plasma into relaxed regions invariant under field-line flow is the only class of ideal-MHD solution that avoids the mathematical pathologies in general 3-D geometries identified by Grad \cite{Grad_67}. 

\begin{figure}[htbp] 
   \centering
		\includegraphics[width = 0.45\textwidth]{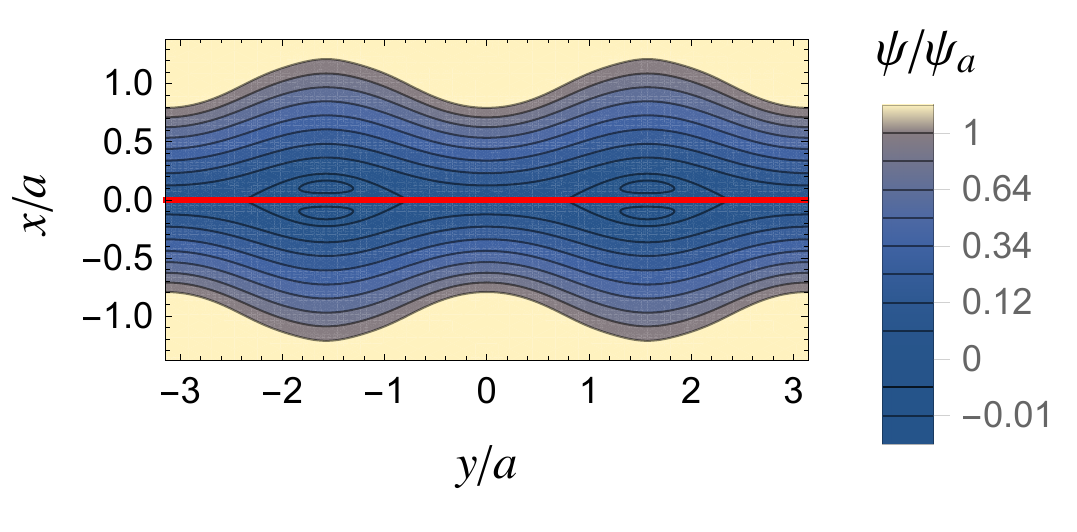} 
\caption{
Boundaries and magnetic surfaces in a typical Hahm--Kulsrud--Taylor (HKT) rippled-boundary case (see Sec.~\ref{sec:rep} for details). The unperturbed boundaries are at $x/a = \pm 1$. The shielding current sheet is along the $y$-axis, which separates the upper and lower relaxed-MHD regions. Note the half islands near the current sheet.
}
\label{fig:Boundary}
\end{figure}

As there are typically local tangential discontinuities across our postulated interfaces 
(and a discontinuity in $|\vrm{B}|$ if they support pressure differences), the interfaces support globally extended \emph{current sheets}. 
On the macroscale on which Taylor relaxation applies, these current sheets are of zero width---their net current may be represented by a Dirac $\delta$ function.
In developing the MRx approach we have normally assumed that 
such long-lived toroidal current sheets 
can exist in general 3-D equilibria only if the rotational transforms $1/q$ on both their inside and outside 
faces are strongly irrational numbers. This assumption is based on a Hamilton--Jacobi construction relating the 
equilibrium surface currents on the two faces combined with Kolmogorov--Arnol'd--Moser (KAM) arguments \cite{Berk_etal_86,McGann_Hudson_Dewar_vonNessi_10,McGann_13}. 
The finite-wavelength stability of such interfaces in general 3-D geometry has yet to be investigated. However, a criterion for ideal-MHD stability to \emph{localized} variations may be established \cite{Barmaz_11} by adapting the energy principle treatment of sharp-boundary equilibria in \cite{Bernstein_etal_58} to show stability when there is no point of zero magnetic shear (no tangential discontinuity) on a surface, suggesting jumps in rotational transform across interfaces are favorable for stability.

Our early development of the multi-region relaxation idea (see e.g. \cite{Hudson_etal_12b}) was based on Taylor's minimum energy variational principle \cite{Taylor_86}, which produces magnetohydro\emph{static} equilibria.
The generalization to a fully fledged fluid dynamics, \emph{Multi-region Relaxed Magnetohydrodynamics} (MRxMHD) has only recently been enunciated \cite{Dewar_Yoshida_Bhattacharjee_Hudson_15}. This new formulation is a \emph{fully dynamical}, time-dependent field theory whose self-consistency is ensured by deriving it from an action principle rather than an energy principle.

The resulting model is simpler and more flexible than ideal MHD, and, we hope, is more physically applicable to fusion physics due to the aforementioned problems with ideal MHD. The existence and stability considerations alluded to above are within the framework of the zero-Larmor-radius, dissipationless MRxMHD formalism itself.  At a minimum, internal consistency of the model justifies using MRxMHD as a regularization and discretization of ideal MHD that is useful for \emph{numerical} purposes. However, to argue that MRxMHD provides a reasonable \emph{physical} model in a specific physical context we need to go beyond its formal framework to consider several questions:
\begin{enumerate}[Q1.\,]\label{en:testQs}
	\item Is there a microscopic mechanism for volume relaxation to occur on a reasonably short timescale? 	\item Is there a mechanism for macroscopic current sheets to form on a similar timescale? 
	\item Are such current sheets robust enough to act as the transport-barrier interfaces postulated in MRxMHD?
	\item Is there experimental evidence that might help answer these questions?
\end{enumerate}

Current interest in modeling RMP penetration using MHD equilibrium codes \cite{Reiman_etal_15,Loizu_etal_15a,Loizu_etal_15b,Loizu_etal_16,Lazerson_etal_16} makes it interesting to explore whether 
dynamical MRxMHD can be used to model the development of shielding current sheets on initially resonant \emph{rational} surfaces as the perturbations are switched on.

In this paper, as well as developing the MRxMHD formalism for boundary-driven time-dependent perturbations, we illustrate it by considering the excitation of a \emph{single} resonant current sheet (to model perhaps the $q = 2$ or 3 surface in Fig.~\ref{fig:DIIID}) by slowly ramping up the amplitude of a sinusoidal rippling of the boundary in  
a very simple geometry, namely the  Hahm--Kulsrud--Taylor (HKT) rippled-boundary slab model \cite{Hahm_Kulsrud_85} illustrated in Fig.~\ref{fig:Boundary}. We address the above questions as plausibly as we are able in this exploratory stage of determining the domain of applicability of the dynamical MRxMHD approach.

The much studied HKT model provides a macroscopic geometry that is computationally simple and has a continuous symmetry in the $z$ direction, thus making the $\vrm{B}$-line dynamics macroscopically integrable (i.e. it has good flux surfaces everywhere). However, we assume, as in DIII-D \cite{Evans_etal_06} (addressing Q1 by appealing to Q4), there are other RMPs simultaneously present that have little effect on the chosen RMP other than the crucial role of providing the microscopic field-line chaos required to justify invoking Taylor relaxation. We assume the only MRxMHD interface is a current sheet forming at the resonant surface $x = 0$, thus dividing the plasma into two equal-pressure relaxation regions, and address Q2 by citing the work of Huang \emph{et al.} \cite{Huang_Bhattacharjee_Boozer_14} who show that current sheets can form rapidly even in the presence of field-line chaos. Even if relaxation applies physically only near $x = 0$, little is lost by assuming
relaxation throughout each plasma subregion, as, in regions with constant equilibrium pressure, adiabatic relaxed MHD agrees with \emph{linearized} ideal-marginal \cite{Newcomb_60} MHD away from resonances \cite{Spies_03,Mills_Hole_Dewar_09}. On the other hand, near the resonant current sheet we find the linear response is weak and nonlinear terms become important, so relaxed MHD is certainly more appropriate than linear ideal MHD \cite{Boozer_Pomphrey_10,White_13} and arguably more appropriate than fully nonlinear ideal MHD \cite{Zhou_Huang_Qin_Bhattacharjee_16}.

While the concept of time is implicit in the present paper, we assume the switching on of boundary ripple 
to be \emph{adiabatic}, i.e. to be sufficiently slow that the system can be considered to evolve through a continuous sequence of steady states. It is our aim to add a quasi-dynamical dimension to the static, equilibrium calculations of Loizu, Hudson \emph{et al.}, \cite{Loizu_etal_15b,Loizu_etal_16} so as to address the physical accessibility of the equilibria with current sheets and discontinuous rotational transform they calculated. This work is also complementary to the ideal-MHD HKT simulation study of Zhou \emph{et al.} \cite{Zhou_Huang_Qin_Bhattacharjee_16}, who demonstrate the formation of a nonlinear ideal current sheet using a variational integrator in Lagrangian labeling that enforces the frozen-in-flux condition exactly. 

By definition, on resonant surfaces $q$ is necessarily rational, but the continuous symmetry of the HKT model gets around the above-mentioned KAM existence argument against rational interface $q$ values because the characteristics of the Hamilton--Jacobi equation are integrable in this case. Furthermore we shall find that, above a small threshold amplitude of the RMP, a discontinuity in $q$ develops across the interface, making $q$ on either side no longer resonant. It has also been shown \cite{Yoshida_Dewar_12,Dewar_Tuen_Hole_17} that an interface located exactly at a rational surface suppresses the associated tearing mode, essentially because the ideal-MHD invariants within the interface suppress reconnection. 

However, in a real plasma, current sheets are neither zero width, nor perfectly conducting---at long times we might expect \cite{Hahm_Kulsrud_85} tearing instability to lead physically to island formation through internal reconnection within the current sheet. If this were so, the answer to Q3 would not be as positive as we would like, as it would reduce the timescale on which MRxMHD applies. But the situation may be rescued by appeal to Q4---in the RMP experiments on DIII-D \cite{Evans_etal_06} there is a strong toroidal flow, and the RMPs have nonzero toroidal mode number. Thus we may invoke the flow-suppression of reconnection discovered by Parker and Dewar \cite{Parker_Dewar_90} in HKT geometry to argue that it is \emph{not} physically unrealistic to assume the interface current sheet to be robust on a long timescale.

In Sec.~\ref{sec:MRxMHD} we summarize the general MRxMHD formalism as presented in \cite{Dewar_Yoshida_Bhattacharjee_Hudson_15}. In addition we discuss the correct relative helicity to use in MRxMHD. We also explain why we can ignore explicit consideration of flow in analyzing the RMP switch-on problem in the HKT model geometry, despite having invoked flow suppression of reconnection above. 

The  HKT model \cite{Hahm_Kulsrud_85} is developed in Sec.~\ref{sec:slab}, where it is shown that MRxMHD leads to a linear Grad--Shafranov equation describing the magnetic field for ripple of arbitrary amplitude. In the HKT model there is reflection symmetry about the resonant current sheet interface, which, in Cartesian coordinates $x,y,z$, we take to be located on the plane $x = 0$. Due to the assumed reflection symmetry, $|\vrm{B}|$ is continuous across $x=0$ so only tangential discontinuities in $\vrm{B}$ can arise. 

In Sec.~\ref{sec:finitemu} we establish the general formalism for calculating solutions of the Beltrami equation in the Grad--Shafranov represention, including, in Sec.~\ref{sec:SlabBeltramiWaves}, a Fourier decomposition of Beltrami fields into plane waves. In \ref{sec:GSWK} we give general  expressions in Grad--Shafranov representation for the magnetic energy, the vector potential and the magnetic helicity.

These formal developments are used to compute the spatially evanescent plasma response to boundary ripple, determining the Fourier coefficients from the boundary conditions, flux and helicity constraints. It is this internal disturbance that resonates at the $x = 0$ magnetic surface to excite the shielding-current-sheet states, which are explored numerically in  \ref{sec:MRxMHDsol} over ranges of initial magnetic shear, ripple amplitude, and poloidal mode number. In Sec.~\ref{sec:Bdy1} studies are performed using amplitudes small enough to need only the lowest spatial harmonic in the Fourier expansion, as in \cite{Hahm_Kulsrud_85}. Scalings with respect to initial magnetic shear and amplitude are determined empirically. It is found that the excited current sheet consists both of a $k_y \neq 0$ ripple response and a net average $k_y = 0$ current. For amplitudes above the threshold value at which the net average current (quadratic in amplitude) begins to dominate the ripple currrent (linear in amplitude), a jump in rotational transform occurs across the current sheet. 

In Sec.~\ref{sec:Bdy2} further studies at higher amplitude are performed by including more terms in the Fourier sums, so as to maintain a sinusoidal boundary ripple. This allows investigation of the dependence on poloidal mode number, $m$, of the threshold amplitude for rotational transform discontinuity. It is found that the threshold is highest at lowest $m$, presumably because the exponential screening of the sheet current ripple is lowest in this case.

Conclusions are given and directions for further work are indicated in Sec.~\ref{sec:conclude}, followed by Appendix~\ref{sec:Kideal} showing why loop integrals $\oint\!\d\vrm{l}\dotv\vrm{A}$ on interfaces must be included as MRxMHD constraint invariants, and Appendix~\ref{sec:vacHel} illustrating why the simple $\int\!\!\vrm{A}\dotv\vrm{B}\,\d V$ form of the magnetic helicity (i.e. with no vacuum helicity subtracted) is the correct helicity constraint invariant to use in MRxMHD under the constraint derived in Appendix~\ref{sec:Kideal}.

An electronic Supplement [Supp] is provided online. It provides further detail on deriving those equations below flagged by a citation to the Supplement. Also in the Supplement is Appendix~\ref{sec:psiBeltrami}, giving a derivation of the Grad--Shafranov form of the Beltrami equation, alternative to that given in Sec.~\ref{sec:rep}, by deriving it directly from the Woltjer--Taylor variational principle of extremizing magnetic energy subject to the constraint of constant magnetic helicity. 

\section{The dynamical MRxMHD model}\label{sec:MRxMHD}

In \cite{Dewar_Yoshida_Bhattacharjee_Hudson_15} the equations for MRxMHD were derived as Euler--Lagrange equations from a  Lagrangian
\begin{equation}\label{eq:LRx}
	L = \sum_i L_i - \int_{\Omega_{\rm v}} \frac{\vrm{B}\dotv\vrm{B}}{2\muSI} \,\d V
\;,
\end{equation}
where the volume integration $\int\! \d V$ in the last term is over a vacuum region $\Omega_{\rm v}$, with $\vrm{B}$ denoting magnetic field and $\muSI$ the permeability of free space.  (However in the HKT model there is no vacuum region, so we do not need this term in the present paper.) The sum $\sum_i$ is over Lagrangians $L_i$ given by
\begin{equation}\label{eq:Li}
	L_i =  \!\int_{\Omega_{i}}\!\!\mathcal{L}^{\rm MHD}\d V + \tau_i(S_i - S_{i0}) +\mu_i\left(K_i - K_{i0}\right) \, .
\end{equation}
Here $\Omega_{i}$ denotes a plasma relaxation region and $\mathcal{L}^{\rm MHD}$ is the standard MHD Lagrangian density \cite{Newcomb_62,Dewar_70}, $\rho v^2/2 - p/(\gamma-1) - B^2/(2\muSI)$, with $\rho$ denoting mass density, $p$ the plasma pressure, and $\gamma$ the ratio of specific heats.

As in ideal MHD, in MRxMHD mass is conserved microscopically (i.e. in each fluid element $\d V$) by constraining it holonomically \cite{Newcomb_62,Dewar_70} to the strain field of Lagrangian fluid element displacements.

However, instead of using ideal-MHD holonomic Lagrangian constraints on $p$ and $\vrm{B}$ to conserve entropy and flux microscopically we treat them as Eulerian fields. The constraint $\divv\vrm{B} = 0$ is enforced by using the representation $\vrm{B} \equiv \curl \vrm{A}$, regarding the vector potential $\vrm{A}$ as an independently variable field, 
which is constrained only by conservation of total magnetic helicity $2\muSI K_i$ in each \emph{macroscopic} subregion $\Omega_i$, and conservation of loop integrals $\oint\!\d\vrm{l}\dotv\vrm{A}$ on the boundaries $\partial\Omega_i$ (see Appendix~\ref{sec:Kideal}). Likewise $p$ is constrained only by conservation of total 
subregion entropies $S_i$. These nonholonomic constraints of constant $K_i$ and $S_i$ are then enforced through Lagrange multipliers $\mu_i(t)$ and $\tau_i(t)$, respectively. 

The entropy $S_i$, given in \cite{Dewar_Yoshida_Bhattacharjee_Hudson_15}, is a functional of $\rho$ and $p$ but its specific form will not be needed in this paper. 
However the magnetic helicity constraint functional,
\begin{equation}\label{eq:Helicity}
	K_i \equiv \int_{\Omega_i} \frac{\vrm{A}\dotv\vrm{B}}{2\muSI} \, \d V \;,
\end{equation}
will play a critical role in our analysis of adiabatic response to ripple switch on. After the Euler--Lagrange equation from variation of $\vrm{A}$ is derived and solved
at each time, $\mu_i$ is chosen so as to satisfy the helicity constraint $K_i - K_{i0} = 0$ (the subtracted constant $K_{i0}$ being the initial value of $K_i$). As the problem we address in this paper involves time-dependent geometric changes in the boundaries we discuss below the constraints on the gauge of $\vrm{A}$ required for $K_i$ to be truly invariant under such boundary changes, an issue not resolved in \cite{Dewar_Yoshida_Bhattacharjee_Hudson_15}. 

We assume the plasma in each relaxation region is entirely enclosed by its boundary (the no-gap condition \cite{Dewar_78a}), implying the tangentiality condition
\begin{equation}\label{eq:Bn}
	\vrm{n}_i\dotv\vrm{B} = 0 \:\: \text{on} \:\: \partial\Omega_i \;,
\end{equation}
where $\partial\Omega_i$ denotes the boundary of region $\Omega_i$ and $\vrm{n}_i$ is the unit normal at each point on $\partial\Omega_i$.

This boundary condition is intimately connected 
with the question of invariance or otherwise of $K_i$ with respect to gauge changes $\vrm{A} \mapsto \vrm{A} + \grad\chi$, which is equivalent to asking whether the volume integral $\int_{\Omega_i}\divv(\vrm{B}\chi)\,\d V$ vanishes or not. Using Gauss' theorem and \eqref{eq:Bn} reduces this volume integral to a sum, over all topologically distinct cross sections $S_l$, of surface integrals $\int_{S_l} \jump{\chi}_l \vrm{n}_l\dotv\vrm{B}\,\d S$, where $\jump{\chi}_l$ denotes the discontinuity (jump) in $\chi$ across $S_l$.

Thus $K_i$ is invariant under variations in $\chi$ if this gauge potential is \emph{single valued}, that is if $\jump{\chi}_l = 0$ over the $\nu$ cross sections $S_l$, where the genus $\nu$ is the number of topologically distinct directions in $\Omega_i$ \cite{Dewar_Yoshida_Bhattacharjee_Hudson_15}. (The genus in our annular tori $\Omega_\pm$ is 2, corresponding to the toroidal and poloidal directions.)

However, single-valued gauge potentials do not exhaust the topologically allowed possibilities: consider transformations of the form $\vrm{A} \mapsto \vrm{A} + \grad\chi^l_{{\rm H}i}$, where the $\nu$ non-single-valued functions $\chi^l_{{\rm H}i}$ are harmonic functions (i.e. solutions of Laplace's equation in $\Omega_i$). The jumps (periods) $\jump{\chi^l_{{\rm H}i}}_l$ are constant over each $S_l$, so $\int_{S_l} \jump{\chi^l_{{\rm H}i}}_l \vrm{n}_l\dotv\vrm{B}\,\d S = \jump{\chi^l_{{\rm H}i}}_l\Phi_l$, where $\Phi_l$ is the magnetic flux through $S_l$. Thus $K_i$ would \emph{not} be gauge invariant with respect to such transformations. However, such transformations during the evolution of the plasma are ruled out by requiring constancy of loop integrals $\oint\!\d\vrm{l}\dotv\vrm{A}$ on the boundaries (the no-gaps condition, see Appendix~\ref{sec:Kideal}), because allowing $\jump{\chi^l_{{\rm H}i}}_l \neq 0$ would change one or more of these loop integrals on $\partial\Omega_i$. 

Most of the loop integrals $\oint\!\d\vrm{l}\dotv\vrm{A}$ can be related to the invariant fluxes $\Phi_l$ within the plasma, but there remains the problem that the magnetic helicities \eqref{eq:Helicity}, while invariant because of the no-gaps condition, are still not uniquely defined because of the \emph{initial} gauge freedom arising from the unknown vacuum poloidal flux threading the toroidal vacuum-plasma interface. There are historically two distinct approaches to fixing this problem, one based on subtracting off products of line integrals $\oint\d\vrm{l}\dotv\vrm{A}$ on boundaries \cite{Taylor_86,Bevir_Gray_82,Spies_03} and the other based on subtracting off corresponding vacuum helicities \cite{Jensen_Chu_84,Berger_Field_84,Finn_Antonsen_85} to form \emph{relative} helicities. 

Both methods involve magnetic fluxes (though represented in different ways) and are both appropriate for \emph{fixed}-boundary problems. However the present problem involves \emph{varying} boundaries and it is not clear that the vacuum helicity is invariant in such cases (see Appendix~\ref{sec:vacHel}), casting doubt on the utility of the relative helicity concept. This quandary is resolved in Appendix~\ref{sec:Kideal} in favor of making invariance of boundary loop integrals $\oint\d\vrm{l}\dotv\vrm{A}$ a fundamental postulate of (no-gaps) MRxMHD but working with the new relative helicity used in \cite{Dewar_Yoshida_Bhattacharjee_Hudson_15}, $K_i - K_{i0}$, which is relative to the \emph{initial} helicity rather than to the \emph{vacuum} helicity.

Variation of $\vrm{A}$ (holding $\mu_i$ fixed) in Hamilton's Action Principle, $\delta\mathscr{S} \equiv \delta\!\int\!L\,\d t = 0$, gives a \emph{Beltrami equation},
\begin{equation}\label{eq:Beltrami}
	\curl\vrm{B} = \mu_i \vrm{B}
\end{equation}
in each subregion
$\Omega_i$, 
to be solved under the tangentiality boundary condition \eqref{eq:Bn}
on $\partial\Omega_i$. 

The action principle also gives the fluid equations within each $\Omega_i$ by varying $p$ and the fluid positions $\bm{\xi}$ under the microscopic mass conservation constraint, which in Eulerian form is 
\begin{equation}\label{eq:rho}
	\frac{\partial\rho}{\partial t} = -\divv(\rho\vrm{v}) \;.
\end{equation}
These variations give the compressible Euler fluid equations for the mass velocity $\vrm{v}$, 
\begin{equation}\label{eq:momentumeq}
	\rho\left(\frac{\partial\vrm{v}}{\partial t} + \vrm{v}\dotv\grad\vrm{v}\right) = -\grad p \;,
\end{equation}
and pressure $p = \tau_i\rho$ \cite{Dewar_Yoshida_Bhattacharjee_Hudson_15}. 

Because of the force-free nature of the magnetic field implied by \eqref{eq:Beltrami}, there is no Lorentz force term in \eqref{eq:momentumeq}---\emph{the plasma flow and magnetic field couple only at the interfaces}. This peculiarity of dynamical MRxMHD makes the HKT geometry particularly attractive: in this geometry the deforming plasma boundaries are externally forced and the interface between the two mirror-image plasma regions is plane, thus allowing us to ignore flow in the subsequent analysis of our simple illustrative case. 

In more general geometries the requirement that sound waves not be excited also sets the slow timescale on which an adiabatic analysis is appropriate. Suffice it to say here that the plasma response to boundary ripple becomes incompressible  in the very low frequency limit \cite{Dewar_Tuen_Hole_17}, so $\rho$ and $p$ are constant in space (and also time if the volumes of $\Omega_i$ are kept constant during ripple switch on).

Variation of fluid positions \emph{at} the interface $\partial\Omega_{i,j} \equiv \partial\Omega_{i}\cap\partial\Omega_{j}$ gives the force-balance condition across the current sheet on this boundary
\begin{equation}\label{eq:normsurfvar}
	\jump{p + \frac{B^2}{2\muSI}}_{i,j} = 0 \;,
\end{equation}
the brackets $\jump{\cdot}_{i,j}$ denoting the jump in a quantity as the observation point crosses the interface from the $\Omega_{i}$ side of to the $\Omega_{j}$ side.

However, due to the reflection symmetry about $x=0$ assumed in our simple HKT-like model, illustrated in Fig.~\ref{fig:Boundary}, the interface between the upper and lower relaxation regions (which we denote by $\Omega_{+}$ and $\Omega_{-}$, respectively) continues to be located on the $x=0$ plane throughout the switching on of the RMP. Also, $B^2$ remains an even function of $x$ and hence continuous (though not necessarily differentiable) across the interface, and also $\jump{p}=0$. Thus \eqref{eq:normsurfvar} is trivially satisfied in this paper.

\section{HKT-Beltrami slab model}\label{sec:slab}
\subsection{Unperturbed pseudo-toroidal equilibrium}\label{eq:toroidalCoords}

In this subsection we limit attention to the initial, unperturbed state of a slab plasma, before boundary ripple is switched on. Then all magnetic field lines can be assumed to lie in parallel planar magnetic surfaces $x = \const$.

To relate slab geometry, as best we can, to that of a toroidal confinement device such as a tokamak, we assume the system to be topologically periodic in $y$ and $z$, with periodic boundary condition lengths $L_{\rm pol} = 2\pi\abdy$ and $L_{\rm tor} = 2\pi R$, respectively. Here $R$ is the nominal major radius of the device and $\abdy$ is a representative radial scale length, typically less than the mean minor radius of an actual plasma. 
The $y$ and $z$ periodic variables are then linearly related to the $2\pi$-periodic poloidal angle $\theta$ and toroidal angle $\zeta$ of a toroidal magnetic coordinate system, 
\begin{equation}\label{eq:yzident}
	\theta = \frac{y}{\abdy} , \quad \zeta = \frac{z}{R} \;. 
\end{equation}

An unperturbed equilibrium field line passing through the point $\theta = \theta_0$,  $\zeta = 0$, on surface $x = x_0$ is then described by the line in $\theta$, $\zeta$ space
\begin{equation}\label{eq:strtfldline}
	\theta = \theta_0 + \iotabar(x_0)\zeta, \quad \zeta = \frac{z}{R} \;, 
\end{equation}
where $\iotabar(x_0)$ [$\equiv 1/q(x_0)$, where $q$ is the unperturbed ``safety factor''] is the \emph{rotational transform} on the flux surface. From \eqref{eq:yzident}, the line \eqref{eq:strtfldline} is given in $x$, $y$ space as
\begin{equation}\label{eq:strtfldlinexy}
	y = \abdy\theta_0 + \iotabar(x)\frac{\abdy z}{R} \;,
\end{equation}
so that the general infinitesimal line element along a field line on an arbitrary surface $x = \mathrm{const}$ is
\begin{equation}\label{eq:linelt}
\begin{split}
	d\mathbf{l} &\equiv dy \esub{y} + dz \esub{z} \\
	& = \left[\iotabar(x)\frac{\abdy}{R}\esub{y} + \esub{z}\right] dz \;,
\end{split}
\end{equation}
by \eqref{eq:strtfldlinexy}, with unit basis vectors $\esub{y} \equiv \grad y$ and $\esub{z} \equiv \grad z$.
Thus the equilibrium magnetic field is parallel to $\iotabar(x)\abdy\esub{y} + R\esub{z}$, so we may write
\begin{equation}\label{eq:Bunpert}
\begin{split}
	\mathbf{B} &=  B(x)\frac{\iotabar(x)\abdy\esub{y} + R\esub{z}}{\sqrt{\iotabar(x)^2\abdy^2 + R^2}}  \\
			    &=  B(x)\frac{\abdy\esub{y} + q(x)R\esub{z}}{\sqrt{\abdy^2 + q(x)^2R^2}}\;. 
\end{split}
\end{equation}
Thus $B_y/B_z = \iotabar \abdy/R$ and $B_z/B_y = qR/\abdy$, giving the well-known expressions
\begin{equation}\label{eq:qunpert}
	\iotabar(x) = \frac{R B_y(x)}{\abdy B_z(x)}\;,\:\: q(x) = \frac{\abdy B_z(x)}{R B_y(x)} \;.
\end{equation}

Assuming an equilibrium with a sheared magnetic field, only an isolated magnetic surface(s) $x = x_{\rm res}$ will \emph{resonate} with a wavelike perturbation with poloidal mode number $m$ such that
\begin{equation}\label{eq:ky}
	k_y \equiv m(2\pi/L_{\rm pol}) = m/\abdy
\end{equation}
and toroidal mode number $n$ [$k_z \equiv -n(2\pi/L_{\rm tor}) = -n/R$] \footnote{We have followed usual toroidal confinement convention by inserting a minus sign in the relation between $k_z$ and $n$.} when the phase fronts coincide with field lines. That is, when $\vrm{k}\cdot\vrm{B} = 0$, which, using \eqref{eq:Bunpert}, is the condition $\iotabar\left(x_{\rm res}\right)m - n = 0$, or
\begin{equation}\label{eq:iotares}
	\iotabar\left(x_{\rm res}\right) = \frac{n}{m}, \quad q\left(x_{\rm res}\right) = \frac{m}{n} \;. 
\end{equation}

In the HKT model \cite{Hahm_Kulsrud_85} $x_{\rm res} = 0$ and the boundary ripple is applied only in the poloidal direction, so $n = 0$, $k_z = 0$, and $k_y = m/\abdy$ \footnote{As experimental RMP coils are not typically $n=0$, we could alternatively regard $y$ and $z$ as helical coordinates rather than poloidal and toroidal coordinates, but the crudeness of a slab model does not seem to warrant such an attempt at further physical realism.}. As $|q\left(x_{\rm res}\right)| = \infty$ we henceforth use only $\iotabar(x)$ to characterize the pitch of the equilibrium field.

We depart from \cite{Hahm_Kulsrud_85} in taking both unperturbed and perturbed magnetic fields to obey \eqref{eq:Beltrami}, though the Lagrange multiplier $\mu$ must change slightly with increasing ripple amplitude in order to satisfy the constant-magnetic-helicity constraint. (However $\mu$ will be the same in both $\Omega_{+}$ and $\Omega_{-}$ due to the assumed symmetry about $x=0$.) Denoting the unperturbed value of $\mu$ by $\mu_0$ (not to be confused with the vacuum permeability $\upmu_0$) we find the unperturbed solution of \eqref{eq:Beltrami},
\begin{equation}\label{eq:BeltramiB0}
	\vrm{B}^{(0)}(x) =  B_0(\sin\mu_0 x\,  \esub{y} + \cos\mu_0 x\, \esub{z}) \;,
\end{equation}
where $B_0$ is a constant. Using \eqref{eq:qunpert} we find the rotational transform in the unperturbed state,
\begin{equation}\label{eq:iotabar0}
	\iotabar(x) = \frac{R}{\abdy} \tan\mu_0 x\;.
\end{equation}

\subsection{Grad--Shafranov (GS) representation}\label{sec:rep}

We follow Hahm and Kulsrud \cite{Hahm_Kulsrud_85} in using a flux-function representation for the full, perturbed magnetic field, defining $\psi(x,y)$ such that
\begin{equation}\label{eq:GSBrep}
	\vrm{B} = F(\psi)\,\esub{z}  + \esub{z} \cross \grad \psi
\end{equation}
where $F(\psi)$ is $B_z$ expressed as a function of $\psi$.

Note that \eqref{eq:GSBrep} implies that $\vrm{B}\dotv\grad\psi \equiv 0$, i.e. $\vrm{B}$ is everywhere tangential to level surfaces $\psi = \const$, so that $\psi$ has the property of being a label for \emph{magnetic surfaces}. Figure~\ref{fig:Boundary} shows rippled magnetic surfaces given by constructing representative contours of $\psi(x,y)$ after $m=2$ [see \eqref{eq:ky}] sinusoidal boundary ripple of amplitude $\alpha = 0.21$ [see \eqref{eq:prescribedBdy}] has been switched on, starting from the ``tokamak-relevant'' case ($\mu_0 \abdy = 0.2$) shown below in Fig.~\ref{fig:eqmPlotsTok}.

Note that $\psi$ is not unique, because it can be changed by a constant amount without changing the observable $\esub{z}\cross\grad\psi$, the \emph{poloidal magnetic field}. Such a baseline shift also changes the functional form of $F$ to retain the invariance of the observable $B_z$, the \emph{toroidal magnetic field}. To remove this arbitrariness we set the baseline for $\psi$ by fixing it on both boundaries $x = \pm\xbdy(y)$ to be the constant value $\psi = \psibdy$: in the HKT model we assume $\psi(x,y)$ to be even in $x$ and to increase away from $x = 0$, as $|y|$ increases, up to $\psibdy$. Though $\psi(x,y)$ is continuous across the current sheet at $x=0$, its derivative $\partial_y\psi$ is in general discontinuous there, so it is sometimes convenient to consider $\psi(x,y)$ as defined on two Riemann sheets intersecting along the cut at $x=0$.

Note that \eqref{eq:GSBrep} implies
\begin{equation}\label{eq:curlB}
	\curl \vrm{B} = \nabla^2\psi\, \esub{z} - F' (\psi)\,  \esub{z}\cross \grad\psi
\end{equation}
so that, crossing \eqref{eq:curlB} with \eqref{eq:GSBrep} ,
\begin{equation}\label{eq:jXB}
	(\curl\vrm{B})\cross\vrm{B} = -(\nabla^2\psi + FF')\grad\psi \;.
\end{equation}
For force-free fields, such as those described by \eqref{eq:Beltrami}, the left-hand side of \eqref{eq:jXB} vanishes, leaving us with the equation
\begin{equation}\label{eq:GSeqn}
	\nabla^2\psi + FF' = 0 \;,
\end{equation}
which is the \emph{Grad--Shafranov} (GS) equation in slab geometry in the special case $p' = 0$. As will be shown below, this is a linear equation and reduces, in the limit $\mu\abdy \to 0$, to the Laplace equation assumed in \cite{Hahm_Kulsrud_85}. The GS representation of Beltrami solutions is also useful in axisymmetric toroidal geometry \cite{Cerfon_ONeil_14}.

Substituting \eqref{eq:GSBrep}  and \eqref{eq:curlB} in \eqref{eq:Beltrami} we get two equations defining a Beltrami field in the GS representation,
\begin{equation}\label{eq:BeltramiGS1}
	\nabla ^2\psi = \mu F
\end{equation}
and
\begin{equation}\label{eq:BeltramiGS2}
	  F'(\psi) = -\mu \;,
\end{equation}
which are consistent with the GS equation, \eqref{eq:GSeqn}. 
Equations (\ref{eq:GSeqn}--\ref{eq:BeltramiGS2}) are also derived variationally from first principles in [Supp]:Appendix~\ref{sec:psiBeltrami} by minimizing magnetic energy at constant helicity. 

Integrating \eqref{eq:BeltramiGS2} gives
\begin{equation}\label{eq:BeltramiF}
	  F(\psi) = C-\mu  \psi
\end{equation}
where $C$ is a spatial constant, though it is not invariant under application of ripple. 
Also, as the left-hand side of \eqref{eq:BeltramiF}, $F = B_z$, is a physical observable, $C$ must counterbalance the arbitrary baseline constant included in $\psi$. 

In the following we find it useful to define the \emph{area-weighted average} of an arbitrary function $f$ over a surface of section across the upper relaxation region $\Omega_{+}$ as
\begin{equation}\label{eq:avdef} 
	\overline{f} \equiv \lbrak f\rbrak \equiv \frac{1}{\Aalpha_{+}}\int_0^{2\pi \abdy}\!\!\!dy\int_{0}^{\xbdy(y)}\!\!\!\!\!\!\!\!dx\, f(x,y) \;,
\end{equation}
where $\Aalpha_{+}$ is the cross-sectional area over one topological periodicity length,
\begin{equation}\label{eq:area}
	\Aalpha_{+} \equiv \int_0^{2\pi \abdy}\!\!\!dy\int_0^{\xbdy(y)}\!\!\!\!\!\!\!\!dx
	 = 2\pi \abdy\lbrak\xbdy\rbrak  \;,
\end{equation}
the total cross-sectional area across the whole plasma being $\Aalpha \equiv \Aalpha_{+} + \Aalpha_{-} = 2\Aalpha_{+}$.
(In general two-relaxation-region problems we would need to define separate averaging operators $\lbrak f\rbrak_{\pm}$ in $\Omega_{+}$ and $\Omega_{-}$, but the reflection symmetry assumed in the HKT model means these averages are equal for even parity functions and the negative of each other for odd parity functions.)
Note we have introduced two equivalent averaging notations, $\lbrak\cdots\rbrak$ being a useful alternative to $\overline{\cdots}$ for lengthy expressions.

To decompose $C$ into an invariant part and a geometrically dependent part we average \eqref{eq:BeltramiF} over a surface of section of $\Omega_{+}$, as in \eqref{eq:avdef}, to give
\begin{equation}\label{eq:Fav}
	C = \Fav + \mu\psiav \;,
\end{equation}
using which \eqref{eq:BeltramiF} becomes
\begin{equation}\label{eq:Ffluct}
	F  = \Fav - \mu\psitld \;,
\end{equation}
where 
\begin{equation}\label{eq:psitildedef}
	\psitld \equiv \psi - \psiav 
\end{equation}
is the deviation from the mean poloidal flux $\psiav$. Determination of the non-invariant quantity $\psiav$ (and hence $C$) will be discussed in Sec.~\ref{sec:GSeqn}.

\begin{figure}[htbp]
\begin{center}
	\includegraphics[width=0.5\textwidth]{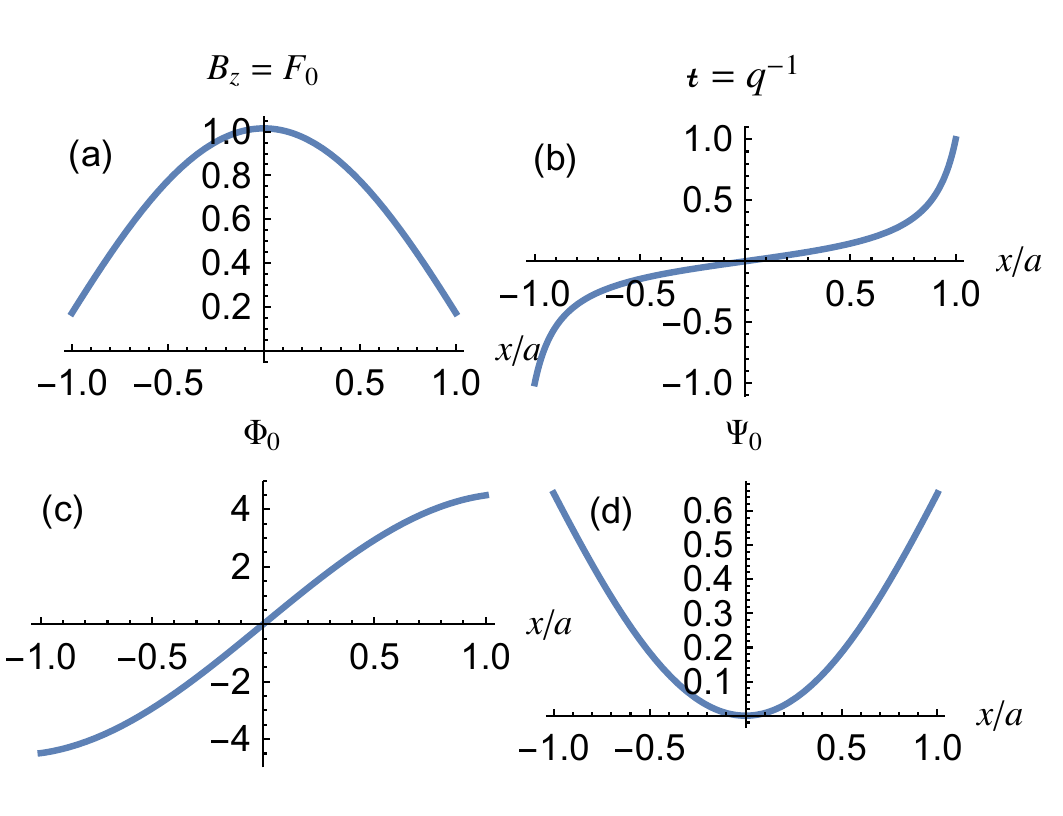}
\caption{Unperturbed (plane boundary) profiles in the RFP-relevant case $\mu_0 \abdy = 1.4$. (a) Toroidal field $B_z = F = B_0\cos\mu_0 x$. (b)  Rotational transform $\iotabar$. c) Toroidal flux $\Phi$. (d) Poloidal flux $\Psi$. Parameters and units are such that $\abdy = B_a = 1$, $R = \abdy \cot\mu_0 \abdy$.}
\label{fig:eqmPlotsRFP}
\end{center}
\end{figure}

\begin{figure}[htbp]
\begin{center}
	\includegraphics[width=0.5\textwidth]{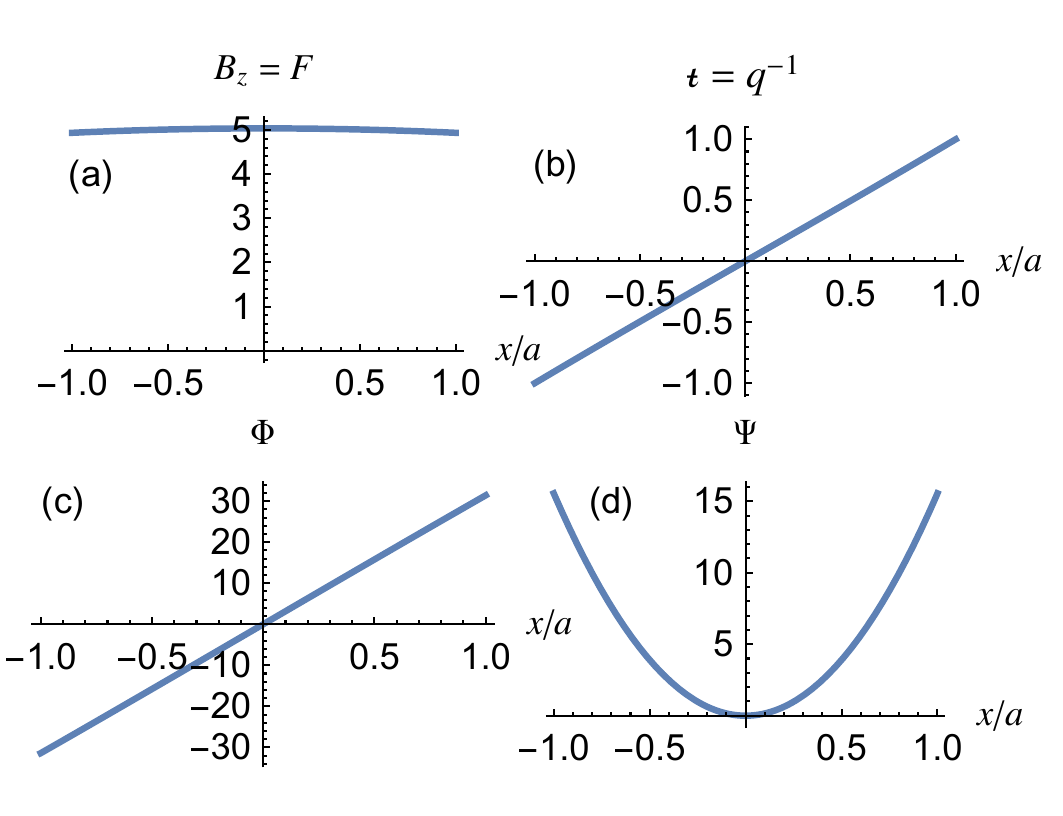}
\caption{Unperturbed (plane boundary) profiles in the tokamak-relevant case $\mu_0 \abdy = 0.2$. (a) Toroidal field $B_z = F = B_0\cos\mu_0 x$. (b)  Rotational transform $\iotabar$. c) Toroidal flux $\Phi^{(0)}(\psi^{(0)}(x))$. (d) Poloidal flux $\Psi$. Parameters and units are as in Fig.~\ref{fig:eqmPlotsRFP}, the choices of $\mu_0$ in the two figures being discussed in Sec.~\ref{sec:unpert}.}
\label{fig:eqmPlotsTok}
\end{center}
\end{figure}

\subsection{Fluxes}\label{sec:flux}

The \emph{poloidal flux} $\Psi(\psi)$ between the current sheet, where $\psi = \psicut$, and a magnetic surface $\psi = \psi_{\rm s}$ is the surface integral of the flux density $\esub{y}\,\dotv\,\esub{z}\cross\grad\psi = \partial_x\psi$ over an area in any plane $y = \const$ bounded by the current sheet $x = 0$,  the magnetic surface labeled by $\psi_{\rm s}$, and the lines $z = \const$ and $z = \const + 2\pi R$,
\begin{equation}\label{eq:polflux}
	\Psi^{\pm}(\psi_{\rm s}) \equiv \int_0^{2\pi R}\!\!\!\!\!\! dz\!\int_{0}^{x^{\pm}(\psi_{\rm s}|y)}\!\frac{\partial\psi}{\partial x}\, dx = 2\pi (\psi - \psicut)R \;,
\end{equation}
where $x = x^{\pm}(\psi_{\rm s}|y)$ denotes the upper ($+$) or lower ($-$) branch of the solution to the equation $\psi(x,y) = \psi_{\rm s}$, for given $y$. (Where $y$ is arbitrary for magnetic surfaces outside half islands such as are seen in Fig.~\ref{fig:Boundary}, but, for defining the ``private flux'' within such an island, $y$ must obviously be restricted to lie within the island.) The linear relation between the poloidal flux $\Psi$ and the function $\psi$ justifies the terminology \emph{poloidal flux function} for the latter. 

The fact that $\Psi(\psi(x,y))$ is an even function of $x$ in the HKT model is illustrated in Figs.~\ref{fig:eqmPlotsRFP} and \ref{fig:eqmPlotsTok} for the unperturbed case \eqref{eq:BeltramiB0} (in which special case there is no current sheet, so $\Psi$ is differentiable at $x=0$).

Assuming here the magnetic surface spans the full poloidal periodicity length $2\pi \abdy$ (i.e. it is not in a half island) we also define the \emph{toroidal flux} $\Phi(\psi)$ as a magnetic surface quantity by integrating the toroidal magnetic field $B_z = F(\psi)$ over one period in $y$ between the resonant surface $x =0$ and the given magnetic surface $x = x^{\pm}(\psi_{\rm s}|y)$ in $\Omega_{\pm}$,
\begin{equation}\label{eq:torfluxfn}
	\Phi^{\pm}(\psi_{\rm s}) \equiv \int_0^{2\pi \abdy}\!\!\!\! dy\!\int_0^{x^{\pm}(\psi_{\rm s}|y)}\! F(\psi_{\rm s}) \, dx \;.
\end{equation}
[Note that $\Phi^{+}(\psi_{\rm s}) = -\Phi^{-}(\psi_{\rm s})$ so $\Phi(\psi(x,y))$ is an odd function of $x$, as illustrated in Figs.~\ref{fig:eqmPlotsRFP} and \ref{fig:eqmPlotsTok}.]
We generalize the ``safety factor'' $q$, defined for the unperturbed field in \eqref{eq:qunpert}, as $q(\psi) \equiv d\Phi/d\Psi$, which, in the GS representation \eqref{eq:GSBrep}, can be written
 \begin{equation}\label{eq:qGS}
	q^{\pm}(\psi_{\rm s}) = \pm\frac{F(\psi_{\rm s})}{2\pi R}\oint_{\rm pol} \frac{dl}{|\grad\psi|}
\end{equation}
where $dl \equiv (dx^2 + dy^2)^{1/2}$ is an element of length along a contour $\psi = \psi_{\rm s}$ running between $y = 0$ and $y = 2\pi \abdy$. This general definition applies equally to the perturbed and unperturbed system.

However, as mentioned in Sec~\ref{sec:slab}, it is more convenient to work with the reciprocal of $q$, the \emph{rotational transform},
\begin{equation}\label{eq:transform}
	\iotabar^{\pm}(\psi) = \frac{d\Psi^{\pm}}{d\Phi^{\pm}}
	 \;.
\end{equation}
As illustrated for the unperturbed case in Figs.~\ref{fig:eqmPlotsRFP} and \ref{fig:eqmPlotsTok}, $\iotabar$ is an odd function of $x$. In this special case it is continuous at $x=0$, but for large enough ripple amplitude we shall find that it may be discontinuous there.

Dotting both sides of \eqref{eq:GSBrep} with $\esub{z}$ and integrating over one wavelength of the cross section, we thus find our first invariant $\Phi_{\rm tor} \equiv \Phi^{+}(\psibdy)-\Phi^{-}(\psibdy) = 2\Phi^{+}(\psibdy)$, the \emph{total toroidal flux}, to be
\begin{equation}\label{eq:torflux}
	\Phi_{\rm tor} = \Aalpha \Fav \;.
\end{equation}
The toroidal flux and magnetic helicity contained between the everywhere perfectly conducting boundaries $x = \pm \xbdy(y)$ are conserved throughout, from switch on to reconnection. Also, to avoid the external work required to change the mean toroidal field $\Fav$ and pressure $p$ we assume the rippling of the walls is done in such  way as to preserve area,
\begin{equation}\label{eq:areaconsn}
	\Aalpha = \Aalpha_0 = 4\pi \abdy^2 \;, 
\end{equation}
which from \eqref{eq:area} is ensured by requiring
\begin{equation}\label{eq:meanxbdy}
	\lbrak \xbdy(y)\rbrak = \abdy \;,
\end{equation}
making the adiabatic plasma response incompressible \cite{Dewar_Tuen_Hole_17}.

With area thus conserved, \emph{toroidal flux conservation is equivalent to invariance of $\Fav$}, which also applies separately in both upper and lower relaxation regions due to the assumed reflection symmetry. Thus in both regions the toroidal flux conservation condition is equivalent to the constraint
\begin{equation}\label{eq:areafluxconsvn}
	\Fav - \Fav_0 = 0
\end{equation}
during switch-on of the boundary ripple perturbation, where $\Fav_0$ denotes the unperturbed value of the mean toroidal field, which is calculated below.

\subsection{Unperturbed state in GS representation}
	\label{sec:unpert}

From \eqref{eq:BeltramiB0} the unperturbed toroidal magnetic field is $F_0(x) \equiv B_0 \cos\mu_0 x$. Thus
\begin{equation}\label{eq:Phi0}
	\Fav_0 = B_0\lbrak\cos\mu_0 x\rbrak_0 \;,
\end{equation}
where
\begin{equation}\label{eq:avcos0}
\begin{split}
	\lbrak\cos\mu_0 x\rbrak_0 &=  \frac{1}{a} \!\!\int_0^\abdy \!\!dx\, \cos\mu_0 x \\
		&= \frac{\sin\mu_0 \abdy}{\mu_0\abdy} \\
		&\sim 1 - \frac{\mu_0^2a^2}{6}+O\left((\mu_0\abdy)^4\right)\;.
\end{split}
\end{equation}

Note the interesting fact that $\Fav_0$ has zeros  at
\begin{equation}\label{eq:BeltramiSpectrum}
	\mu_0\abdy = \pi n
\end{equation}
for integer $n \neq 0$, corresponding to extreme reversed-field states.  However we shall not consider such large values of $\mu_0\abdy$ in this paper.

For use later in this paper it will be found useful to define $\vrm{U}$, the plane-slab unit-vector solution of \eqref{eq:Beltrami} with general $\mu$,
\begin{equation}\label{eq:Beltrami0}
	\vrm{U}(x|\mu) \equiv  \sin\mu x  \,\esub{y} + \cos\mu x \,\esub{z} \;.
\end{equation}
In terms of $\vrm{U}$, the unperturbed field $\vrm{B}^{(0)}(x)$, given above by \eqref{eq:BeltramiB0}, can be represented as $B_0\vrm{U}(x|\mu_0)$.

In the GS representation, $\vrm{U}$ can be represented by
\begin{subequations}
\begin{align}
	 \psiU(x|\mu) &\equiv \frac{1}{\mu}(1 - \cos\mu x) = \frac{2}{\mu}\sin^2 \frac{\mu x}{2}\;, \label{eq:psiU}\\
	F_{\rm U}(x|\mu) &\equiv \cos \mu x = 1 - \mu\psiU \label{eq:FU}\;.
\end{align}
\end{subequations}
Setting $\psi^{(0)}(x) = B_0\psiU(x|\mu_0)$, $F_0(\psi^{(0)}) = B_0 F_{\rm U}(x|\mu_0)$ in \eqref{eq:GSBrep} verifies that $\vrm{B}^{(0)}(x) = B_0\vrm{U}(x|\mu_0)$.
Note that we have chosen the arbitrary constant in the sheared-field flux function to be such that $\psiU = 0$ on the $y$-axis. Note also the useful identity
\begin{equation}\label{eq:psiUidentity}
	\psiU^{\prime 2} + \mu^2\psiU^2 = 2\mu\psiU \;,
\end{equation}
where $\psiU' \equiv \partial\psiU/\partial x$.

For devices such as the reversed-field pinch (RFP), $\mu_0 \abdy$ is $O(1)$ \cite{Taylor_86}. Such a case, close to the value $\pi/2$ where the toroidal field changes sign at the boundary, is plotted above in Fig.~\ref{fig:eqmPlotsRFP}.
However this RFP-like example is not strongly relevant to devices like tokamaks and stellarators. For these devices $B$ is dominated by the toroidal magnetic field $B_z = F(\psi)$, which can be modeled in MRxMHD by taking $\mu_0 \abdy \ll 1$. Although large, $B_z$ is then approximately constant, the interesting physics being in the behavior of the \emph{poloidal} field $B_y = \partial_x\psi$. Thus, rather than specifying $B_0$ directly, we find it more convenient to specify the \emph{boundary poloidal field}, $B_a = \psi'_a \equiv B_0\psi'_{\rm U}(a|\mu_0)$.
Using \eqref{eq:psiU} $B_0$ is then given by
\begin{equation}\label{eq:B0}
	 B_0 =  \frac{B_a}{\sin \mu_0 \abdy} \;,
\end{equation}
which diverges in the limit $\mu_0 \abdy \to 0$.

We denote the boundary value of the poloidal flux function by $\psibdy \equiv B_0\psiU(a|\mu_0)$. Using Eqs.~(\ref{eq:psiU}) and \ref{eq:B0}, this is given by [Supp] 
\begin{equation}\label{eq:psiaconstraint}
	\psibdy = \frac{B_a}{\mu_0}\tan\frac{\mu_0 \abdy}{2} \;,
\end{equation}
which approaches $\half \abdy B_a$ as $\mu_0 \abdy \to 0$.

To develop a tokamak-relevant set of parameters, we set the unperturbed boundary rotational transforms to be $\iotabar^{(0)} = \pm 1$ at $x = \pm\abdy$. Then \eqref{eq:iotabar0} gives $\abdy/R = \tan\mu_0 \abdy \approx \mu_0 \abdy$. For our \emph{standard tokamak-relevant reference case} we take $\mu_0 \abdy = 1/5$,  $\abdy/R =\tan\mu_0 \abdy \approx 0.2027$, giving the aspect ratio $R/\abdy\approx 5$ as shown earlier in Fig.~\ref{fig:eqmPlotsTok}.

\section{Rippled states in GS representation}
\label{sec:finitemu}

\subsection{Rippled boundary conditions}
	\label{sec:init}
The perfectly-conducting boundary walls are then deformed (rippled) by switching on, over a time short compared with the reconnection timescale for the resulting long-lived current sheet, wavelike perturbations with (fundamental) poloidal wave number $k_y$.
Reflection symmetry of both walls and plasma about the $y$-axis is assumed, so $\psi$ remains an even function of $x$.

From \eqref{eq:Beltrami0}, $\vrm{U}(0) = \esub{z}$ so the resonance condition $\vrm{k} \dotv\vrm{B} = 0$ (see Sec.~\ref{sec:SlabBeltramiWaves}) is satisfied at $x = 0$, i.e. along the $y$-axis, where a shielding current sheet of full width $2\pi\abdy$ initially forms to prevent island formation. This current sheet cuts $\Omega$ into two disjoint subdomains, $\Omega_{+}$, between $x = 0$ and upper boundary $x = \xbdy(y)$, and $\Omega_{-}$, between $x = 0$ and the lower boundary $x = -\xbdy(y)$.

We shall find the \emph{fully shielded state}, immediately after the boundary perturbation is switched on, by assuming Taylor relaxation occurs independently in $\Omega_{\pm}$ [a special case of the Multi-Region Relaxed MHD (MRxMHD) problem \cite{Hudson_etal_12b}], so the perturbed initial magnetic fields in $\Omega_{\pm}$, before reconnection of the shielding current sets in, are found by solving \eqref{eq:Beltrami} under the boundary and other conditions discussed below.

On an equilibrium current sheet it can be shown \cite[Appendix A]{McGann_Hudson_Dewar_vonNessi_10} that the normal component of $\vrm{B}$ must vanish. In terms of the representation \eqref{eq:GSBrep}, $\psi$ is thus constant on both sides of the current sheet. Also $\psi$ must be \emph{continuous} across the current sheet.

By definition, in the fully shielded state no poloidal flux has yet been reconnected through $x = 0$. Also, no poloidal flux can escape through the perfectly conducting walls. Thus the current sheet boundary condition $\psi = \psicut$ at $x=0$ applies, with $\psicut$ fixed at its unperturbed value, which we have chosen to be
\begin{equation}\label{eq:polfluxcons}
	\psicut  \equiv 0 \;.
\end{equation}
On the rippled walls the flux function remains $\psi = \psibdy$ to conserve poloidal flux.
Likewise toroidal flux is trapped between the walls and current sheet, consistently with  \eqref{eq:areafluxconsvn}. 

We consider two methods for defining the boundary waveform function $\xbdy(y)$:
\begin{enumerate}[{Bdy-}1.]
\item\label{itm:HKTbdy}
	The indirect \emph{implicit boundary method} \cite{Dewar_Bhattacharjee_Kulsrud_Wright_13} where we specify the boundary conditions on $\psi$
\begin{equation}\label{eq:HKTbs}
	\psi(\pm \abdy,y) - \lbrak\psi(\pm \abdy,y)\rbrak
	 = 2\alpha\, \psibdy \cos  k_y y \;.
\end{equation}
The factor $\psibdy$, defined in \eqref{eq:psiaconstraint}, is introduced in \eqref{eq:HKTbs} to make the \emph{ripple amplitude parameter} $\alpha$ dimensionless, the factor 2 being to make $\alpha$ the same as in Ref.~\onlinecite{Dewar_Bhattacharjee_Kulsrud_Wright_13} in the limit $\mu_0 \abdy \to 0$.

The function $\xbdy(y)$ is then defined by the contour $\psi = \psibdy \; : \; x = \xbdy(y|\alpha)$, with $\psi$ constructed so as to enforce \eqref{eq:polfluxcons}, the area/toroidal flux conservation equation \eqref{eq:areaconsn}, and the magnetic helicity constraint $K_i - K_{i0} = 0$. The domain $\Omega = \Omega_{-}\!\cup\Omega_{+}$ is now completely specified, its complete boundary $\partial\Omega$ being the union of the two external boundaries $x = \pm \xbdy(y)$ and the internal boundary formed by the cut along the $y$-axis.  Note that we do not linearize with respect to $\alpha$ so $\xbdy(y)$ is not an exact sinusoid in this method. 
\item\label{itm:sinbdy}
	The direct \emph{explicit boundary method}, as used in Fig.~\ref{fig:Boundary}, where we prescribe $x_{\rm bdy}$ to be exactly sinusoidal,
\begin{equation}\label{eq:prescribedBdy}
	x_{\rm bdy}(y) = \abdy(1 - \alpha\cos k_y y) \;.
\end{equation}
\end{enumerate}
The advantage of method Bdy-\ref{itm:HKTbdy} is that it allows a simple closed-form solution of the perturbed Beltrami equation similar to the type found by Hahm and Kulsrud \cite{Hahm_Kulsrud_85}. The disadvantage is that $\xbdy(y)$ becomes highly non-sinusoidal at quite moderate ripple amplitudes $\alpha$ and the method breaks down as $\alpha$ increases further. 

Method Bdy-\ref{itm:sinbdy}, on the other hand, can treat large ripple amplitudes, as evidenced for instance in Fig.~\ref{fig:Boundary}. Its disadvantage is that $\psi$ must be expanded in an infinite series of higher harmonics if \eqref{eq:prescribedBdy} is to be satisfied exactly [see \eqref{eq:psigen}]. However, in practice a good approximation can be found with a reasonable number of expansion functions. For small $\alpha$ the two methods are equivalent.

\subsection{GS equation boundary conditions}
\label{sec:GSeqn}

In the present application of the GS formulation our simple boundary condition on the current sheet, \eqref{eq:polfluxcons} allows us to identify $C$ immediately as $F(0)$, the toroidal magnetic field on the current sheet at $x=0$. However, for $\alpha \neq 0$, $F(0)$ is not known \emph{a priori} but must be determined along with $\psi$ in the solution procedure.

Unlike $F(0)$, $\Fav$ is a known constant, from \eqref{eq:areafluxconsvn}. Also $\psitld$ is independent of the arbitrary constant in $\psi$ and is thus a better flux variable to work with. Like $\psi$ it must be constant on the boundaries and current sheets, obeying the  boundary conditions
\begin{subequations}
\begin{align}
	 \psitld\left(\xbdy(y),y\right) & = \psibdy - \psiav \;, \quad\:\:
	 			 \forall\: y  \label{eq:psitldbdy}\\ 
	 \psitld(0,y) & = \psicut - \psiav \;, 
	 					\label{eq:psitldcut}
\end{align}
\end{subequations}

Unlike $\psi$, neither of these boundary values is known \emph{a priori}. Instead $\psiav(\alpha)$ 
needs to be determined, along with $\mu$, under the Taylor relaxation constraints and the constraint implied by \eqref{eq:psitildedef},
\begin{equation}\label{eq:psitildeav}
	\lbrak\psitld\rbrak \equiv 0 \;.
\end{equation}

Substituting \eqref{eq:Ffluct} in \eqref{eq:BeltramiGS1} gives a \emph{linear} GS equation in the form of an inhomogeneous Helmholtz equation,
\begin{equation}\label{eq:linGSeqn}
	 (\nabla^2 +\mu^2)\psitld = \mu \Fav \;.
\end{equation}
Averaging both sides of \eqref{eq:linGSeqn} and using \eqref{eq:psitildeav} we find $\lbrak\nabla^2\psitld\rbrak = \mu \Fav$ with
\begin{equation}\label{eq:del2av}
	\lbrak\nabla^2\psitld\rbrak = -\frac{1}{\Aalpha_{+}}\int_{\partial\Omega_{+}} \vrm{n}\dotv\grad\psitld \,\d l \;,
\end{equation}
where the right-hand side is found by applying Gauss' theorem, with $\d l = (\d x^2 + \d y^2)^{1/2}$ an element of length along two contours, the upper wall $x = \xbdy(y)$ and the upper side of the current sheet $x=0$ between $y  = -\pi\abdy$ and $y = \pi\abdy$, $\vrm{n}$ being the \emph{inward directed} unit normal at each point on $\partial\Omega_{+}$.

Noting from \eqref{eq:psiU} that $(\nabla^2 +\mu^2)\psiU = \mu$, solving \eqref{eq:linGSeqn} can be reduced to the solution of a homogeneous equation using the ansatz
\begin{equation}\label{eq:separansatz}
	 \psitld(x,y) = \Fav\psiU(x|\mu) + \psihat(x,y) \;.
\end{equation}
where $\psihat$ obeys the \emph{homogeneous Helmholtz equation}
\begin{equation}\label{eq:HelmholtzB}
	 (\nabla ^2 + \mu ^2)\psihat = 0
\end{equation}
under the boundary and averaging conditions following from Eqs.~(\ref{eq:psitldbdy} - \ref{eq:psitildeav})
\begin{subequations}
\begin{align}
	 \psihat(\xbdy(y),y) & = \psibdy - \psiav  - \Fav\psiU(\xbdy(y)|\mu) \:\:
	 			 \forall\: y \;, \label{eq:psihatbdy}\\ 
	 \psihat(0,y) & = \psicut - \psiav \quad\quad
	 			 \forall\: y \in \text{cuts} \;, \label{eq:psihatcut}\\ 
	 \lbrak{\psihat}\rbrak & =  - \Fav\lbrak \psiU\rbrak \label{eq:psihatav} \;. 
\end{align}
\end{subequations}

The parameter $\lbrak\psi_{\rm U}\rbrak$ is a functional of the boundary shape, which may or may not be known \emph{a priori} depending on whether we use method Bdy-\ref{itm:HKTbdy} or Bdy-\ref{itm:sinbdy}. The unknown  parameters to be solved for are $\mu$, $\psiav$, and coefficients of terms in the ansatz for $\psihat(x,y)$ to be discussed in Sec.~\ref{sec:SlabBeltramiWaves}.

\subsection{Unperturbed state: include only $k_y = 0$}
\label{sec:LaminarBeltrami}

In this section we calculate expressions for initial (unperturbed) states before ripple is switched on. 
The unperturbed state is defined by $\alpha = 0$, with only $x$-dependent magnetic field $\vrm{B}^{(0)}(x) = B_0\vrm{U}(x|\mu_0)$, where $\vrm{U}$ is given by \eqref{eq:Beltrami0}. The corresponding flux function is 
\begin{equation}\label{eq:psi0}
	\psi_0(x) \equiv \psi^{(0)}(x) \equiv B_0\psiU(x,\mu_0)  \;,
\end{equation}
where $\psiU$ is defined in \eqref{eq:psiU}.

Applying the averaging operator defined in \eqref{eq:avdef}, with unperturbed boundary, we find 
[Supp] 
\begin{equation}\label{eq:psi0av}
\begin{split}
	\psiav_0 
		&= \frac{B_0}{\mu_0}\left(1 - \cosav\right) \\
\end{split}
\end{equation}
where $\cosav$ is defined in \eqref{eq:avcos0}.
Hence, in \eqref{eq:psitildedef},
\begin{equation}\label{eq:psi0tilde}
\begin{split}
	\widetilde{\psi}_0(x) 
		&\equiv B_0\psiU(x|\mu_0) - \psiav_0 \\
		&= \frac{B_0}{\mu_0}\left(\frac{\sin\mu_0 \abdy}{\mu_0 \abdy} - \cos\mu_0 x\right)\\
		&= \frac{a B_0}{2}\left[\left(\frac{x^2}{a^2} - \frac{1}{3}\right)\mu_0 \abdy + O\left(\mu_0^3\right)\right]\;.
\end{split}
\end{equation}

The decomposition \eqref{eq:separansatz}, $\psitld_0(x) = \Fav_0\psiU(x,\mu_0) + \psihat_0(x)$, implies
\begin{equation}\label{eq:psi0hat}
\begin{split}
	\psihat_0(x)
	&\equiv \psitld_0(x) - \Fav_0\psiU(x,\mu_0) \\
	&= \frac{\Fav_0 - B_0}{\mu_0}\cos\mu_0 x \\
	&= \frac{B_0}{\mu_0}\left(\frac{\sin\mu_0 \abdy}{\mu_0 \abdy} - 1\right)\cos\mu_0 x \;,
\end{split}
\end{equation}
which is in the kernel of $\nabla^2 + \mu^2$ as required by \eqref{eq:HelmholtzB}.

\subsection{Rippled state: include $k_y \neq 0$ terms}
\label{sec:SlabBeltramiWaves}

Here we generalize the Hahm--Kulsrud \cite{Hahm_Kulsrud_85} solutions by expanding in a basis of plane-wave Beltrami solutions---a $k_y = 0$ solution $B\vrm{U}(x|\mu)$, with $B$ and $\mu$ to be determined, and ``ripple'' solutions that are periodic in the $y$ direction and exponential in the $x$ direction.

The general solution of the Beltrami equation \eqref{eq:Beltrami} is a superposition of divergence-free plane wave solutions with wave vector $\vrm{k}' = \pm k'_x\esub{x} \pm k'_y\esub{y}$ such that $\vrm{k}^{\prime 2} = \mu^2$. To satisfy $2\pi\abdy$ topological periodicity in the $y$ direction we introduce the \emph{poloidal mode number}, $m' = 0, 1, 2, \ldots$, such that $k'_y \equiv m'/\abdy$. Thus \eqref{eq:Beltrami} implies
\begin{equation}\label{eq:kxcond}
	k_x^{\prime 2} \equiv \mu^2 - k_y^{\prime 2}  = \mu^2 - \frac{m^{\prime 2}}{\abdy^2} \;.
\end{equation}
The $y$-independent solutions considered in the previous section correspond to $m' = 0$, giving $k'_x = \pm\mu$. Ripple solutions of Hahm--Kulsrud type require \emph{imaginary} $k'_x$, so we consider only the case $|\mu| < k'_y$ and set $k'_x = \pm i\kappa_m(\mu)$, where
\begin{equation}\label{eq:kmudef}
	\kappa_{m'}(\mu) = (k_y^{\prime 2} - \mu^2)^{1/2} \equiv \left(\frac{m^{\prime 2}}{\abdy^2} - \mu^2\right)^{1/2} \;.
\end{equation}

In the above, $m'$ is the fundamental poloidal mode number $m$ of the imposed ripple or a harmonic, $m' = l m$, where $l = 0, 1, 2, 3, \ldots$. We denote the fundamental ripple wavelength by
\begin{equation}\label{eq:lambdam}
	\lambda_m = \frac{2\pi\abdy}{m} \;.
\end{equation}

The requirement $\divv\vrm{B} = 0$ is ensured by using the $F,\psi$ representation \eqref{eq:GSBrep}, the most general $2\pi\abdy$-periodic solution of \eqref{eq:HelmholtzB}, analytic on the halfplane $x > 0$, being 
\begin{equation}\label{eq:psigen}
\begin{split}
	\psihat_{+}(x,y) &= c_0\cos\mu x
	+ \sum_{l=1}^{\infty} c_{l m} \cos \frac{l m y}{\abdy}\, \cosh \kappa_{l m} x \\
	&\:\: + d_0\sin\mu x + \sum_{l=1}^{\infty} d_{l m} \cos \frac{l m y}{\abdy}\, \sinh \kappa_{l m} x \;.
\end{split}
\end{equation}
with the corresponding solution $\psihat_{-}(x,y)$ on the halfplane $x < 0$ being given by the symmetry condition $\psihat_{-}(x,y) = \psihat_{+}(|x|,y)$.

Our generalized Hahm--Kulsrud-type solutions are superpositions of the form \eqref{eq:separansatz}, $\psi = \psiav + \Fav\psiU + \psihat$ on the cut $x,y$-plane, with the branch $\psihat = \psihat_{\pm}$ being chosen according as $x \gtrless 0$ and the constants $\{c\}$, $\{d\}$, $\psiav$, $\psicut$ and $\mu$ being chosen as described qualitatively in Secs.~\ref{sec:init}. Method Bdy-1 uses only $l = 0$ and $l = 1$ while Method Bdy-2 in principle uses $l = 0,\dots,\infty$.

\section{Energy and Helicity in GS Representation}\label{sec:GSWK}

\subsection{Relative magnetic energy density}
\label{sec:energy}

Rather than use the total magnetic energy $W = \int_\Omega B^2/2\upmu_0 \,\d V$ we find it neater to define $\cal{W}$, the average energy per unit volume, multiplied by $\muSI$. That is, $\mathcal{W}$ is defined as $\lbrak B^2\rbrak/2$, which we henceforth refer to simply as the energy density, there being no thermal energy, as we have taken $p = 0$, and kinetic energy being negligible in adiabatic processes. Although we defined the averaging operation in \eqref{eq:avdef} to be over $\Omega_{+}$, the same results apply in $\Omega_{-}$ due the assumed symmetry and the fact that the energy (and helicity---see below) densities are even functions of $x$.

In GS representation \eqref{eq:GSBrep}, and using \eqref{eq:Ffluct}, the energy density is given by
\begin{equation}\label{eq:W}
\begin{split}
	\mathcal{W} &= \frac{1}{2}\lbrak F^2 + |\grad\psi|^2 \rbrak\\
	 &= \frac{\Fav^2}{2}  + \WSigma \;,
\end{split}
\end{equation}
where we have defined $\WSigma$, the energy density in the internally generated (i.e. non-vacuum) field, as, using integration by parts, [Supp] 
\begin{equation}\label{eq:WSigmaBel}
\begin{split}
	\WSigma
	&= \frac{1}{2}\lbrak |\grad\psitld|^2 + \mu^2\psitld^2\rbrak \\
	&= \frac{1}{2}(\psibdy-\psiav)\mu\Fav + \frac{\psibdy - \psi_{\rm cut}}{2a\lambda_m} J_{+} + \mu^2\lbrak\psitld^2\rbrak
	 \;,
\end{split}
\end{equation}
the  ripple wavelength $\lambda_m$ being defined by \eqref{eq:lambdam}.
In the above manipulations we have used Eqs~\ref{eq:psitildedef}--\ref{eq:del2av} and the boundary conditions Eqs.~\ref{eq:psitldbdy} and \ref{eq:psitldcut} and have denoted the line integrals on the top/bottom surface of the current sheet cut over one ripple wavelength by $J_{\pm}$,
\begin{equation}\label{eq:Jdef}
	J_{\pm} \equiv \pm\int_{-\lambda_m/2}^{\lambda_m/2}\psitld_x(0\pm,y) \, dy
\end{equation}
where $\psitld_x(x,y) \equiv \partial_x\psitld(x,y)$. Taking into account the symmetry about the $y$-axis we have $J_{-} = J_{+}$. 

To discover the physical meaning of $J_{\pm}$ note that the strength of a sheet current is $j_{\ast} = \jump{\vrm{n}_{+}\dotv\grad\psitld}/\muSI$, where $\vrm{n}_{+}$ is the unit normal at each point on the upper surface of the current sheet and $\jump{\cdot}$ denotes the jump in this normal direction. In our case the current sheet is at $x = 0$ and $\vrm{n}_{+} = \esub{x}$, so
\begin{equation}\label{eq:sheetcurr}
	j_{\ast}(y) = \frac{1}{\muSI}[\psitld_x(0+,y) - \psitld_x(0-,y)] \;.
\end{equation}
Integrating \eqref{eq:sheetcurr} along the current sheet we see that the \emph{total current} per ripple period in the current sheet is $J/\muSI$ where $J \equiv J_{+} + J_{-} = 2J_{+}$.

Using \eqref{eq:psi0tilde} in \eqref{eq:WSigmaBel} we find the unperturbed internal magnetic energy,
\begin{equation}\label{eq:WSigmaBel0}
\begin{split}
	\mathcal{W}_{\Sigma 0}
	&= \frac{B_0^2}{2} \left[1 - \cosav^2\right] \\
	&\sim \frac{B_a^2}{6}[1  + O(\abdy^2\mu_0^2)] \;,
\end{split}
\end{equation} 
The latter form being found by using \eqref{eq:avcos0}, eliminating $B_0$ using \eqref{eq:B0}, and expanding in $\mu_0 \abdy$. 

From \eqref{eq:W} we see that, in the ``tokamak-relevant'' small-$\mu_0\abdy$ limit (see Sec.~\ref{sec:unpert}), $\mathcal{W}/\mathcal{W}_{\Sigma 0}$ is dominated by the large vacuum toroidal field energy term $\Fav^2/2\mathcal{W}_{\Sigma 0}$, which, from \eqref{eq:B0}, is seen to diverge like $1/\mu_0^2\abdy^2$ in this limit. However $\Fav$ is invariant under application of ripple because of our constant-volume constraint \eqref{eq:areaconsn}. Thus it is more instructive to work with the \emph{relative energy density}, 
\begin{equation}\label{eq:RelEnergy}
	\Delta\mathcal{W} = \WSigma - \mathcal{W}_{\Sigma 0} \equiv \Delta\WSigma \;,
\end{equation}
where the vacuum toroidal field energy has cancelled out.

\subsection{Relative magnetic helicity}\label{sec:helicity}

It is readily verified, by calculating $\vrm{B} = \curl\vrm{A}$ and comparing with \eqref{eq:GSBrep}, that
\begin{equation}\label{eq:GSArx}
	\vrm{A} =  -\psi\esub{z} + \frac{1}{\mu }\esub{z}\cross\grad\psitld 
\end{equation}
is a vector potential satisfying the requirement (see Appendix~\ref{sec:Kideal}) of invariance of the loop integrals $\oint_{\rm pol}\d\vrm{l}\dotv\vrm{A} = 2\pi\abdy\psicut$ and $\oint_{\rm tor}\d\vrm{l}\dotv\vrm{A} = 2\pi R\psicut$ on the current sheet (topologically a torus). 

By eliminating the linear term in $\psi$ from \eqref{eq:GSArx} using \eqref{eq:BeltramiF}, an alternative form,
\begin{equation}\label{eq:GSAlt}
\begin{split}
	\vrm{A}
	 	&= \frac{\vrm{B} - C\esub{z}}{\mu} \\
\end{split}
\end{equation}
is found that will be useful below for relating energy and helicity. 

By analogy with $\mathcal{W}$ in Sec.~\ref{sec:energy} we define the \emph{average helicity density} $\mathcal{K} \equiv \lbrak \vrm{A}\dotv\vrm{B}\rbrak/2$ and the \emph{relative helicity density} $\Delta\mathcal{K} \equiv \lbrak \vrm{A}\dotv\vrm{B}\rbrak\!/2 - \lbrak \vrm{A}\dotv\vrm{B}\rbrak_0\!/2$, so that, comparing with \eqref{eq:Helicity} the helicity constraint $K_i - K_{i0} = 0$ is equivalent to
\begin{equation}\label{eq:Kconstrt}
	\Delta\mathcal{K} = 0 \;.
\end{equation}

From \eqref{eq:GSAlt}, then \eqref{eq:W} and $\esub{z}\dotv\vrm{B} = F$, and eliminating $C$ with \eqref{eq:Fav}, we find the general expression [Supp] 
\begin{equation}\label{eq:Hel}
\begin{split}
	\mathcal{K} &\equiv \frac{\lbrak \vrm{A}\dotv\vrm{B}\rbrak}{2}  \\
		&= \frac{\WSigma}{\mu} - \frac{\psiav\,\Fav}{2}  \;,
\end{split}
\end{equation}
expressing a linear relation between helicity and energy. The importance of the constant offset in this relation. arising in general from a surface term, $\int_{\partial\Omega}\vrm{A}\cross\vrm{B}\dotv\vrm{n}\,\d S$, in discussing minimum energy states was originally pointed out by Reiman \cite{Reiman_80,Reiman_81}. 
Below we use this result to obtain the analytical version for $\mathcal{K}$ that is used in the numerical studies presented in Sec.~\ref{sec:MRxMHDsol}.  As $\mathcal{K} = \mathcal{K}_ 0$ after the unknowns are solved for numerically, \eqref{eq:Hel} is used again to calculate the numerical value of $\WSigma$.

The unperturbed helicity density $\mathcal{K}$ is simply a special case of \eqref{eq:Hel}. Subtracting it from $\mathcal{K}$ and using the invariance of $\Fav$ and \eqref{eq:WSigmaBel} gives
\begin{equation}\label{eq:RelHel}
\begin{split}
	\Delta\mathcal{K}
		&=  -(\psiav-\psiav_0)\Fav + \frac{\psibdy - \psi_{\rm cut}}{2a\mu\lambda_m} J_{+}  \\ 
		& \quad + \mu\lbrak\psitld^2\rbrak - \mu_0\lbrak\psitld_0^2\rbrak
		 \;.
\end{split}
\end{equation}

\section{Shielded RMP solutions}\label{sec:MRxMHDsol}

To explore the properties of this model quantitatively we find Beltrami solutions satisfying rippled boundary conditions, appropriate to the methods discussed in Sec.~\ref{sec:init}, and the boundary conditions Eqs.~(\ref{eq:psihatcut}--\ref{eq:psihatav}) on the current sheet. To provide a complete set of equations to solve numerically we also impose helicity conservation, \eqref{eq:Kconstrt}. As \eqref{eq:meanxbdy} conserves cross-sectional area, toroidal flux conservation is equivalent to the conservation of $\Fav$, which is thus still given by \eqref{eq:Phi0}.

\subsection{HKT-like rippled boundary condition}\label{sec:Bdy1}

\begin{figure}[htbp]
   \centering
		\includegraphics[width = 0.45\textwidth]{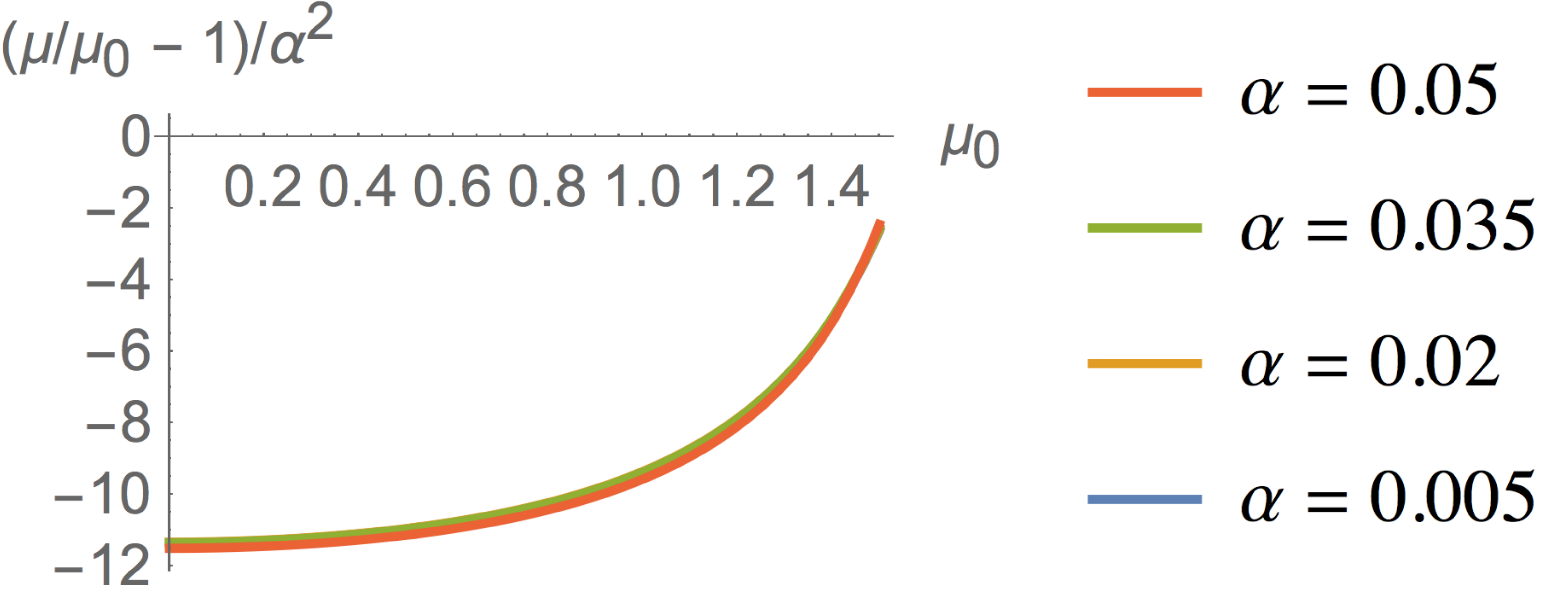} 
\caption{Bdy-\ref{itm:HKTbdy}, $\lambda = a =m =1$: Plots of $[\mu(\alpha,\mu_0)-\mu_0]/(\mu_0\alpha^2)$ \emph{vs}. $\mu_0$ for selected values of $\alpha$ in the range $[0,0.05]$. The coincidence of the curves shows the $\alpha$-dependence has been scaled out to high accuracy. }
\label{fig:mush}
\end{figure}

In this subsection we use the method Bdy-\ref{itm:HKTbdy} to specify the rippled boundaries. Units such that $2\pi/k_y = a =1$ have been used, as well as the choice $\lambda \equiv 2\pi/k_y = a$. In this subsection we also assume $L_{\rm pol} = a$, so $m=1$, but in the standard convention used elsewhere in this paper, $L_{\rm pol} = 2\pi a$, this would be equivalent to taking $m = 2\pi \approx 6$.  

The analog of the Hahm--Kulsrud solution with a shielding current sheet on the resonant surface $x=0$ (denoted in \cite{Hahm_Kulsrud_85} by subscript I, here denoted by subscript ``sh''), is the special case of \eqref{eq:psigen}
\begin{equation}\label{eq:pishatShielded}
\begin{split}
	\psihat_{\rm sh}(x,y) 
		&\equiv \frac{2\alpha\psibdy}{\sinh (\kappa_m \abdy)}\left(|\sinh \kappa_m x|\cos k_y y 
		\phantom{\frac{\kappa_m}{\mu}}\right. \\
	 	&\quad + \left. \gammas\frac{\kappa_m}{\mu} |\sin\mu x|\right)
			 - \psiav\cos\mu x \;,
\end{split}
\end{equation}
where, from \eqref{eq:kmudef}, $\kappa_m(\mu) = (m^2/\abdy^2 - \mu^2)^{1/2}$, where $\mu = \mu_{\rm sh}(\alpha,\mu_0)$, is to be determined. Comparing with \eqref{eq:psigen}, we have set $c_1$ and all $l>1$ coefficients to zero, but have kept all other $l=0$ and $l=1$ terms. The coefficient $d_1 = 2\alpha\psibdy/\sinh (\kappa_m \abdy)$ has been chosen so that $\psi(\abdy,y)$ automatically satisfies \eqref{eq:HKTbs}, the parameter $\alpha$ setting the ripple amplitude. Also the coefficient $c_0 = -\psiav$ has been chosen so that $\psi(0,y)$ automatically satisfies \eqref{eq:psihatcut}, with the poloidal flux conservation condition \eqref{eq:polfluxcons}.

\begin{figure}[htbp]
   \centering
		\includegraphics[width = 0.45\textwidth]{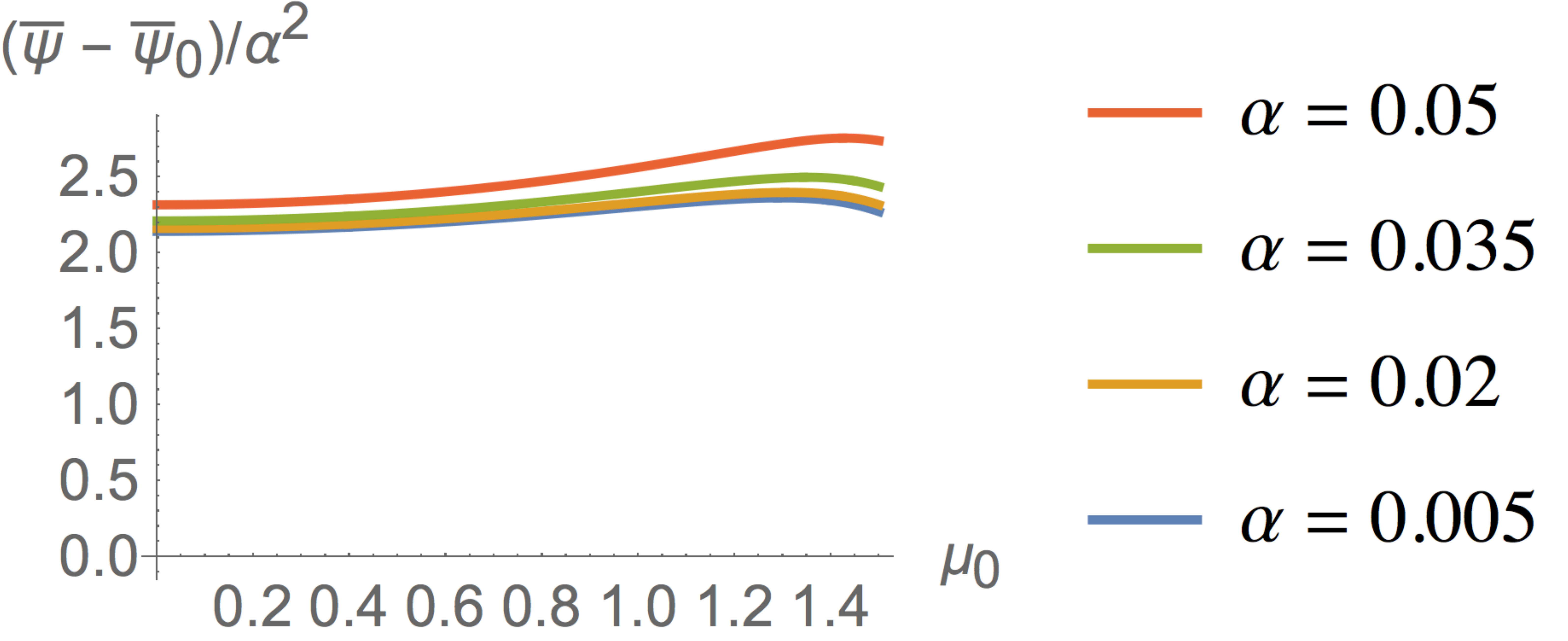} 
\caption{Bdy-\ref {itm:HKTbdy},  $\lambda = a = m =1$: Plots of $[\psiav(\alpha,\mu_0) - \psiav(0,\mu_0)]/\alpha^2$ (in units such that $B_a = 1$) \emph{vs}. $\mu_0$ for selected values of $\alpha$ in the range $[0,0.05]$. (In these plots the vertical ordering of $\alpha$ in the legends is the same as that for the corresponding curves, shown online by color.)}
\label{fig:psiavsh}
\end{figure}

\begin{figure}[htbp]
   \centering
		\includegraphics[width = 0.45\textwidth]{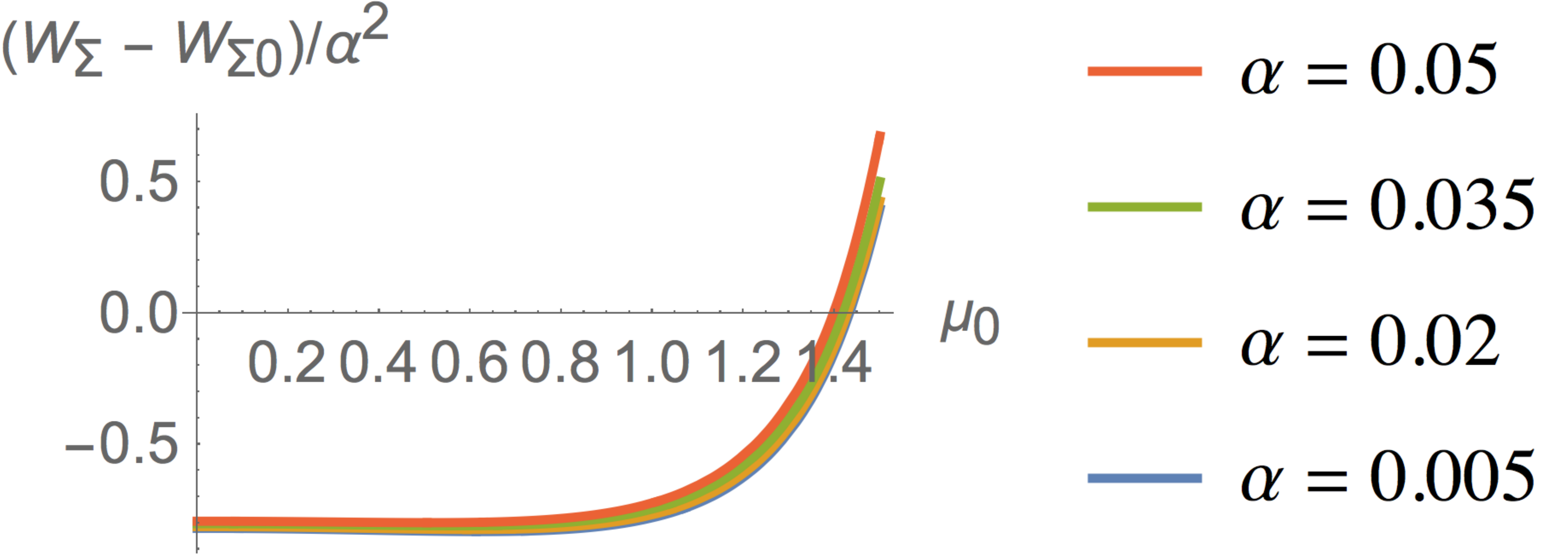} 
\caption{Bdy-\ref {itm:HKTbdy},  $\lambda = a = m =1$: Plots of $\Delta\WSigma(\alpha,\mu_0)/\alpha^2$ \emph{vs}. $\mu_0$ for selected values of $\alpha$ in the range $[0,0.05]$. }
\label{fig:WSigmash}
\end{figure}

\begin{figure}[htbp]
   \centering
		\includegraphics[width = 0.45\textwidth]{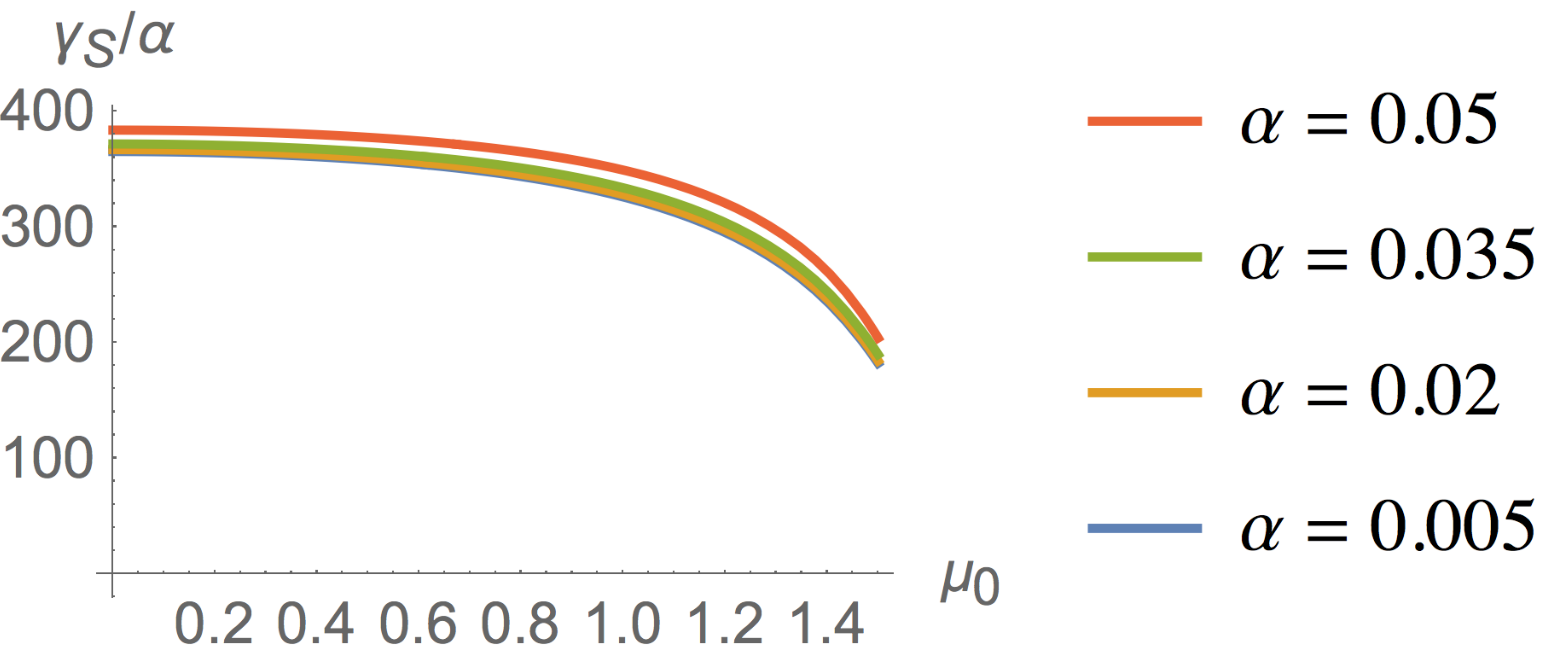} 
\caption{Bdy-\ref {itm:HKTbdy},  $\lambda = a = m =1$: Plots of $\gamma_{\rm S}(\alpha,\mu_0)/\alpha$ \emph{vs}. $\mu_0$ for selected values of $\alpha$ in the range $[0,0.05]$.}
\label{fig:gammaSsh}
\end{figure}

By analogy with Eq.~(13) of Ref.~\onlinecite{Dewar_Bhattacharjee_Kulsrud_Wright_13}, in our expression for $\psihat_{\rm sh}$ above we have renormalized the amplitude $d_0$ of the $\sin\mu x$ term by setting $d_0 = 2\alpha\psibdy \gammas  \kappa_m/\mu\sinh (\kappa_m \abdy)$, the dimensionless parameter $\gammas$ adding a constant term to the sheet current on the $x$-axis---from \eqref{eq:pishatShielded}, the jump in $\partial_x\psitld$ in the expression, \eqref{eq:sheetcurr}, for the sheet current $j_{\ast}$ is given by
\begin{equation}\label{eq:sheetstrength}
	\jump{\partial_x\psihat_{\rm sh}} \equiv  \frac{4\alpha\psibdy \kappa_m  }{\sinh \kappa_m  \abdy}(\cos k_y y + \gammas)  \;,
\end{equation}
so that the total-current parameter $J$, defined below \eqref{eq:sheetcurr}, becomes
\begin{equation}\label{eq:Jsh}
	J =  \frac{4\alpha\psibdy \kappa_m \lambda_m }{\sinh \kappa_m  \abdy}\,\gammas \;.
\end{equation}

The boundary function $\xbdy^{\rm sh}(y|\alpha)$ is determined from \eqref{eq:psihatbdy},  the three parameters $\mu = \mu_{\rm sh}(\alpha,\mu_0)$, $\psiav = \psiav_{\rm sh}(\alpha,\mu_0)$, and $\gammas = \gammas^{\rm sh}(\alpha,\mu_0)$ being determined by solving the 3 simultaneous equations \eqref{eq:meanxbdy}, \eqref{eq:psitildeav} and \eqref{eq:Kconstrt}, with $\mathcal{K}$ given by \eqref{eq:RelHel}. The average energy density can then be found from \eqref{eq:WSigmaBel} or \eqref{eq:Hel}.
 
\begin{figure}[htbp]
   \centering
		\includegraphics[width = 0.45\textwidth]{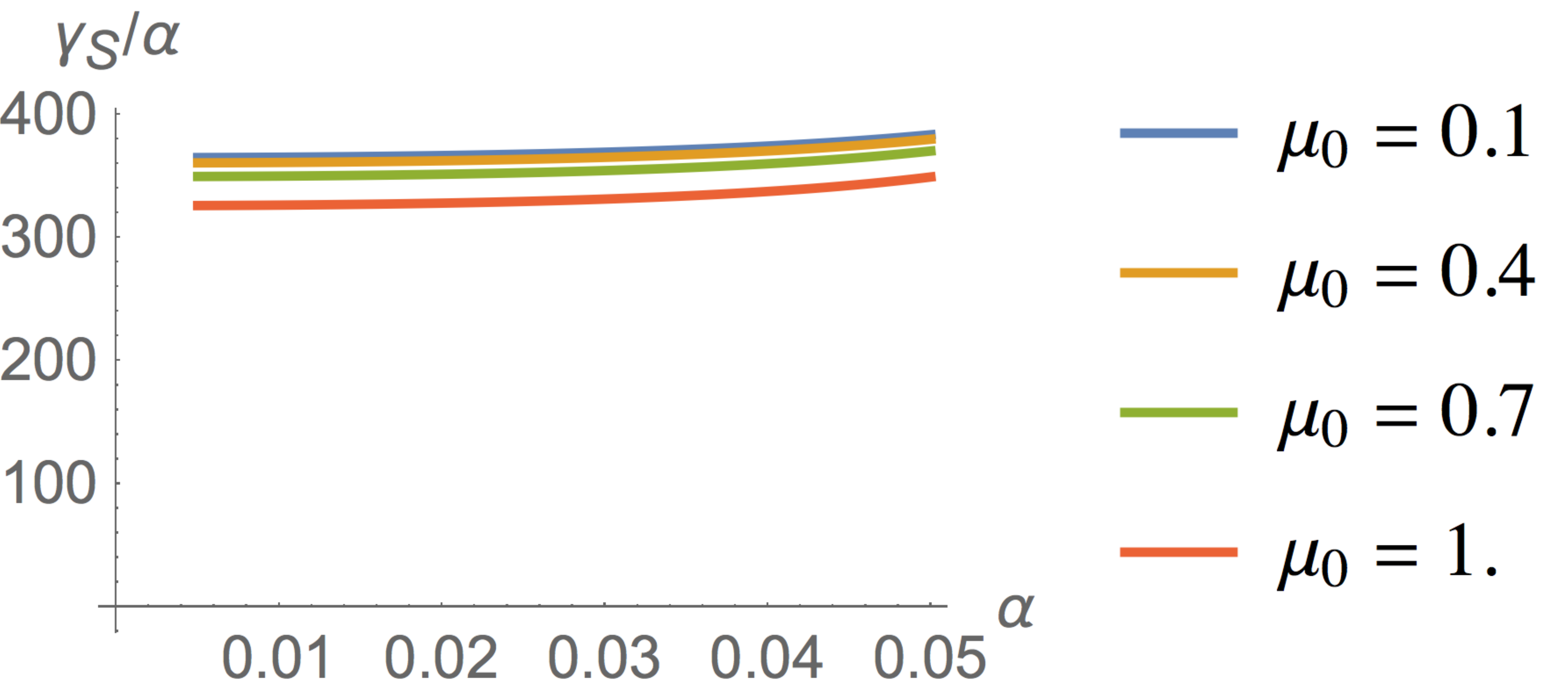} 
\caption{Bdy-\ref {itm:HKTbdy},  $\lambda = a = m =1$: Plots of $\gamma_{\rm S}(\alpha,\mu_0)/\alpha$ \emph{vs}. $\alpha$ for selected values of $\mu_0\abdy$ within the restricted range $[0,1]$, showing $\gamma_{\rm S}/\alpha$ is approximately constant with respect to both variables in these ranges.}
\label{fig:gammaSshalpha}
\end{figure}

Numerical results showing the $\mu_0\abdy$-dependence of $\mu$, $\psiav$, $\WSigma $, and $\gammas$ in the case $k_y = 2\pi/\abdy$, also used in Ref.~\onlinecite{Dewar_Bhattacharjee_Kulsrud_Wright_13}, in units such that $\abdy = 1, B_a = 1$ [see \eqref{eq:B0}], are given in Figs. \ref{fig:mush}--\ref{fig:gammaSshalpha}. The fact that the scaled curves for different values of $\alpha$ are almost identical show that the small-amplitude scalings $\mu-1 \propto \alpha^2$, $\psiav - \psiav_0 \propto \alpha^2$, $\WSigma  - \mathcal{W}_{\Sigma 0} \propto \alpha^2$, and $\gammas  \propto \alpha$ are a good approximation for the range $\alpha < 0.05$ depicted (becoming exact in the limit $\alpha \to 0$). For $\mu_0 \abdy < 1$ the plotted quantities are approximately constant with respect to $\mu_0$, but vary more rapidly above this range as $\mu_0 \abdy$ approaches the value $\pi/2 \approx 1.57$ at which $B^{(0)}_z = B_0\cos \mu_0 x$ reverses sign at $x = \pm \abdy$ [cf. Fig.~\ref{fig:eqmPlotsRFP}(a).].

Quite apart from demonstrating a mathematical scaling law, the physics shown in Fig.~\ref{fig:WSigmash} is worthy of remark because of the sign of $\Delta\WSigma$ and its reversal at large values of $\mu_0\abdy$. First note from \eqref{eq:RelEnergy} that $\Delta\WSigma$ and $\Delta\mathcal{W}$ are equal, so the negative values of $\Delta\WSigma$ at small-to-moderate $\mu_0\abdy$ means the total magnetic field energy in the plasma \emph{decreases} as ripple is imposed. That is, the plasma does work on the boundary. It is tempting to interpret this as implying that a slab plasma confined by MRxMHD interface current sheets at the vacuum-plasma boundaries would be unstable toward spontaneous rippling. 

However, this conclusion would be unwarranted as the total energy includes not only the $O(\alpha^2)$ \emph{wave} energy in the $m \neq 0$ ripple, but also small, $O(\alpha^2)$, nonlinear corrections to the energy in the $m = 0$ \emph{background} state. The sign of such a background energy correction is dependent on the precise nature of the rippling process as it could easily be changed by allowing an $O(\alpha^2)$ change in the cross-sectional area $\Aalpha$, rather than arbitrarily imposing its constancy through through the constraint \eqref{eq:areaconsn}. Furthermore, proper stability analysis of a free-boundary MRxMHD plasma \cite{Hole_Hudson_Dewar_07,Dewar_Tuen_Hole_17} must include the change in the vacuum energy outside the plasma. Such issues will be discussed further elsewhere.

\begin{figure}[htbp]
   \centering
		\includegraphics[width = 0.45\textwidth]{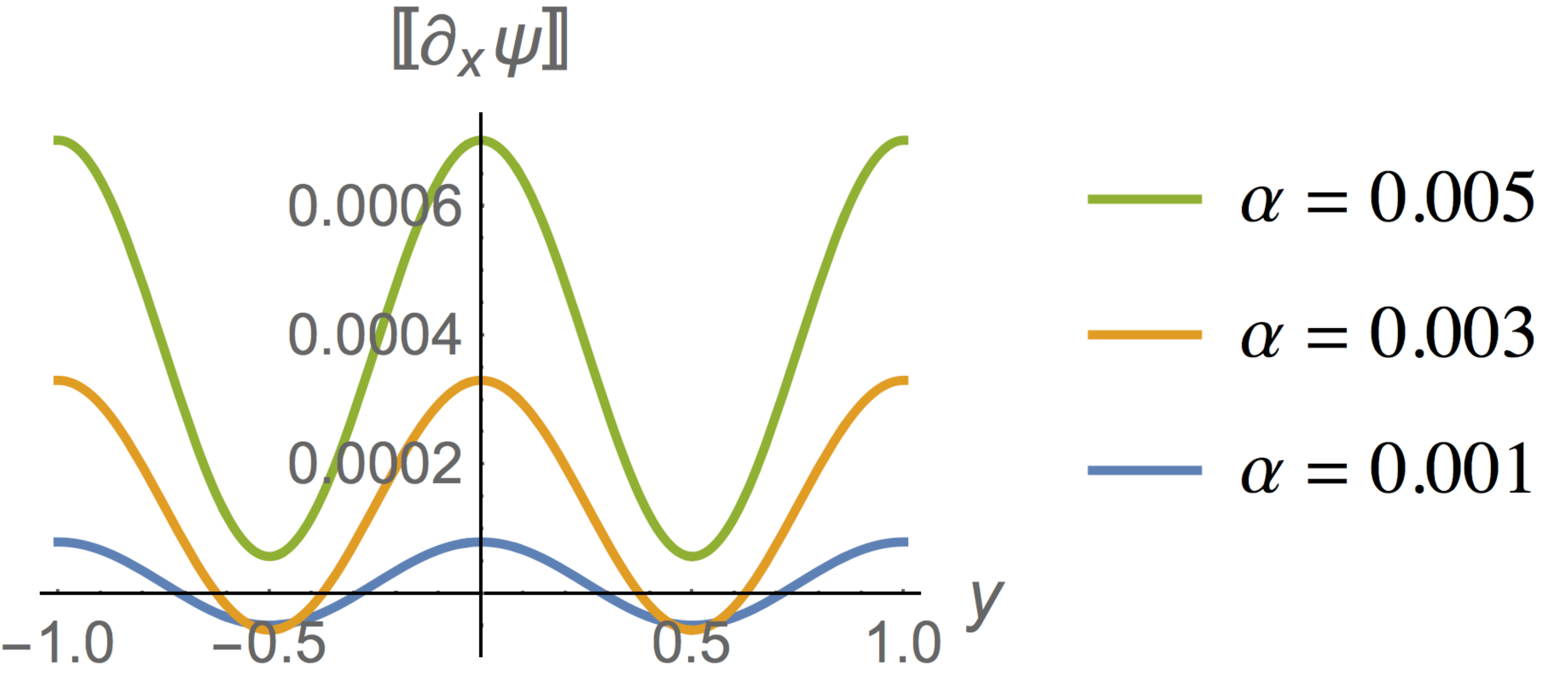} 
\caption{Bdy-\ref {itm:HKTbdy},  $\lambda = a = m =1$: Plots of the jump in the gradient of $\psi$, \eqref{eq:sheetstrength}, \emph{vs}. $y/\lambda$ for $\mu_0 = 1.4/\abdy$ and selected small values of $\alpha$, showing the occurrence of current-density reversal for the two smallest values. }
\label{fig:sheetcurrent}
\end{figure}

The linear $\alpha$-dependence of the dimensionless parameter $\gammas$, shown more explicitly in Fig.~\ref{fig:gammaSshalpha}, is particularly interesting, in the light of \eqref{eq:sheetstrength}, as it means the $m = 0$ response $J$ scales as $\alpha^2$, i.e. it is \emph{nonlinear}. Thus, for small-enough values of $\alpha$, the $m = 0$ response (the term in $\gammas$) is dominated by the linear response (the term in $\cos k_y y$). In this case the singular current density reverses sign over a range of $y$. However, as shown in Fig.~\ref{fig:sheetcurrent}, above the very small threshold value, $\alpha_{\rm thr}$, at which $\gammas = 1$, the $m = 0$ nonlinear response becomes increasingly dominant and the sheet current becomes of constant sign. (Both figures use the same parameters, $\abdy = \lambda_1 = 1$, as the previous plots.)

\begin{figure}[htbp]
   \centering
		\includegraphics[width = 0.45\textwidth]{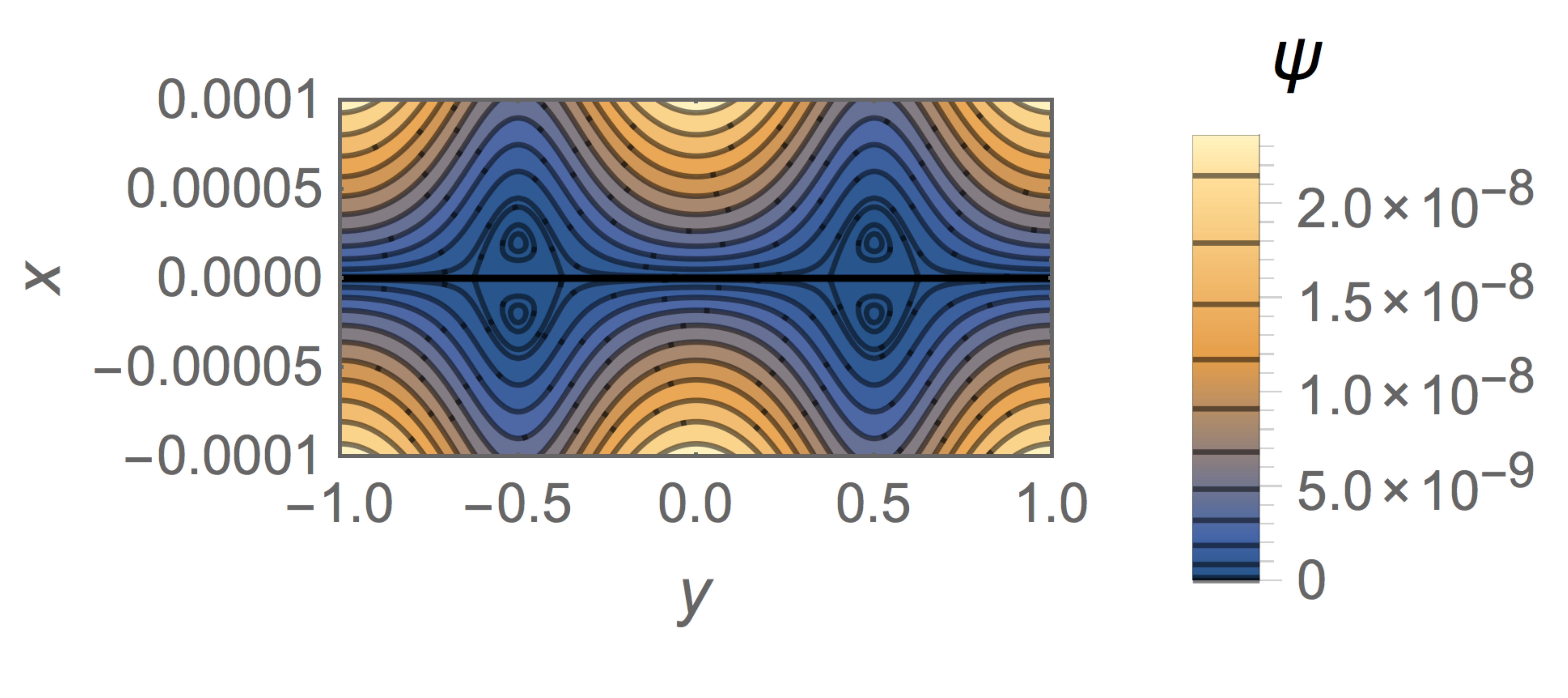} 
\caption{Bdy-\ref {itm:HKTbdy},  $\lambda = a = m =1$: Level surfaces of $\psi$ (magnetic surfaces) in the case $\mu_0 = 1.4/\abdy$, $\alpha = 0.003 < \alpha_{\rm thr}$, showing pairs of a small islands separated by the reversed-current section of the current sheet along the $x$-axis shown in Fig.~\ref{fig:sheetcurrent}. }
\label{fig:sheetsurfs2}
\end{figure}

\begin{figure}[htbp]
   \centering
		\includegraphics[width = 0.45\textwidth]{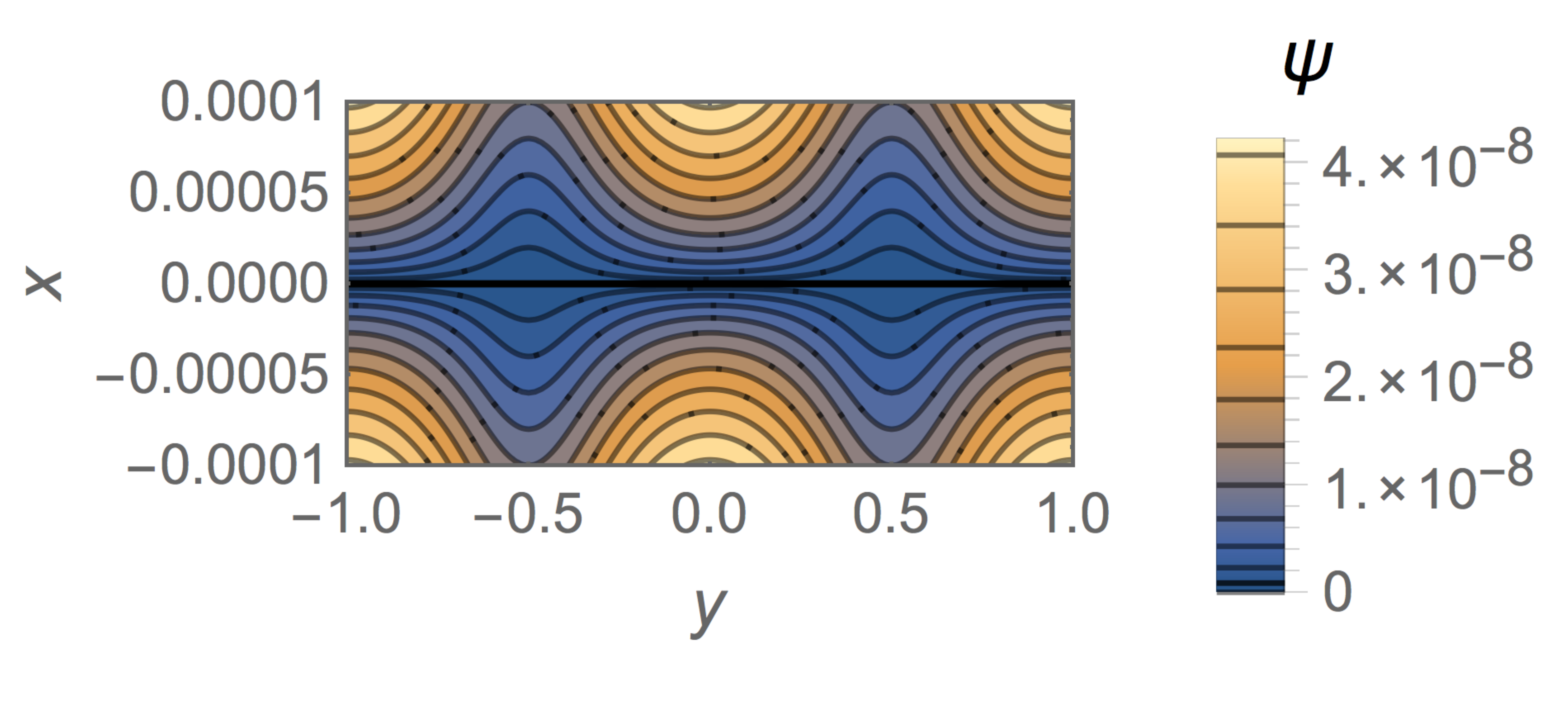} 
\caption{Bdy-\ref {itm:HKTbdy},  $\lambda = a = m =1$: Level surfaces of $\psi$ in the case $\mu_0 = 1.4/\abdy$, $\alpha = 0.005 > \alpha_{\rm thr}$, for which Fig.~\ref{fig:sheetcurrent} shows there is no current reversal and hence no magnetic islands. }
\label{fig:sheetsurfs3}
\end{figure}

\begin{figure}[htbp]
   \centering
		\includegraphics[width = 0.45\textwidth]{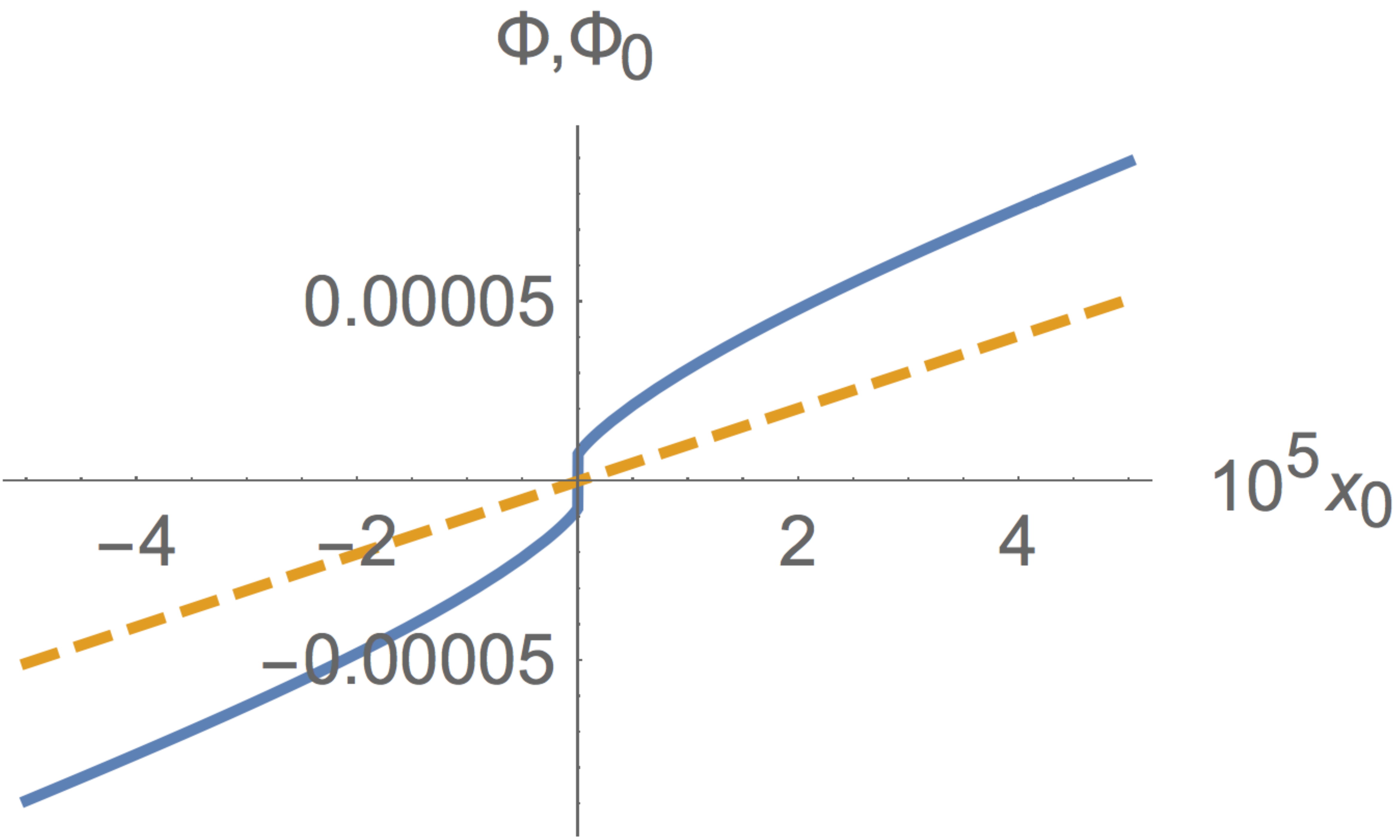} 
\caption{Bdy-\ref {itm:HKTbdy},  $\lambda = a = m =1$, taking $R = \abdy$: Toroidal flux \emph{vs.} the magnetic surface label $x_0$, the $x$-value where a $\psi$-contour crosses the $x$-axis, in the case $\mu_0 = 1.4/\abdy$, $\alpha = 0.003 < \alpha_{\rm thr}$. The dashed curve is for the unperturbed case $\alpha = 0$. }
\label{fig:TorFlx_003}
\end{figure}
As the poloidal field at the current sheet is proportional to the current-sheet strength, \eqref{eq:sheetcurr},  such a transition has a profound effect on the topology of the magnetic surfaces close to the current sheet, as illustrated in Figs. 5 and 6 of Ref.~\onlinecite{Dewar_Bhattacharjee_Kulsrud_Wright_13} and Figs~\ref{fig:sheetsurfs2} and \ref{fig:sheetsurfs3} of the present paper.  It is seen that small islands form near the current sheet in the current-reversal case $\alpha < \alpha_{\rm thr}$
\footnote{This effect occurs in the linear case (i.e. when $\alpha^2$ terms are negligible), as is also observed in linearized ideal MHD, where it is inconsistent with the frozen-in flux condition \cite{Boozer_Pomphrey_10,White_13}. However, in relaxed MHD the linearity of the Beltrami equation is exact, and there is no inconsistency as topology change is allowed.},
causing the contour $\psi = \psicut = 0$ to trifurcate into upper and lower magnetic surfaces $x = x^{\pm}_{\psi}(y|\psicut)$ and the resonant surface $x = 0$. In this case the toroidal flux function $\Phi$, \eqref{eq:torfluxfn}, jumps  at $x = 0$ by the amount of ``private'' toroidal flux in the islands, as shown in Fig.~\ref{fig:TorFlx_003}. For $\alpha > \alpha_{\rm thr}$, $\Phi(x_0)$ is continuous at $x_0$.

\begin{figure}[htbp]
   \centering
		\includegraphics[width = 0.45\textwidth]{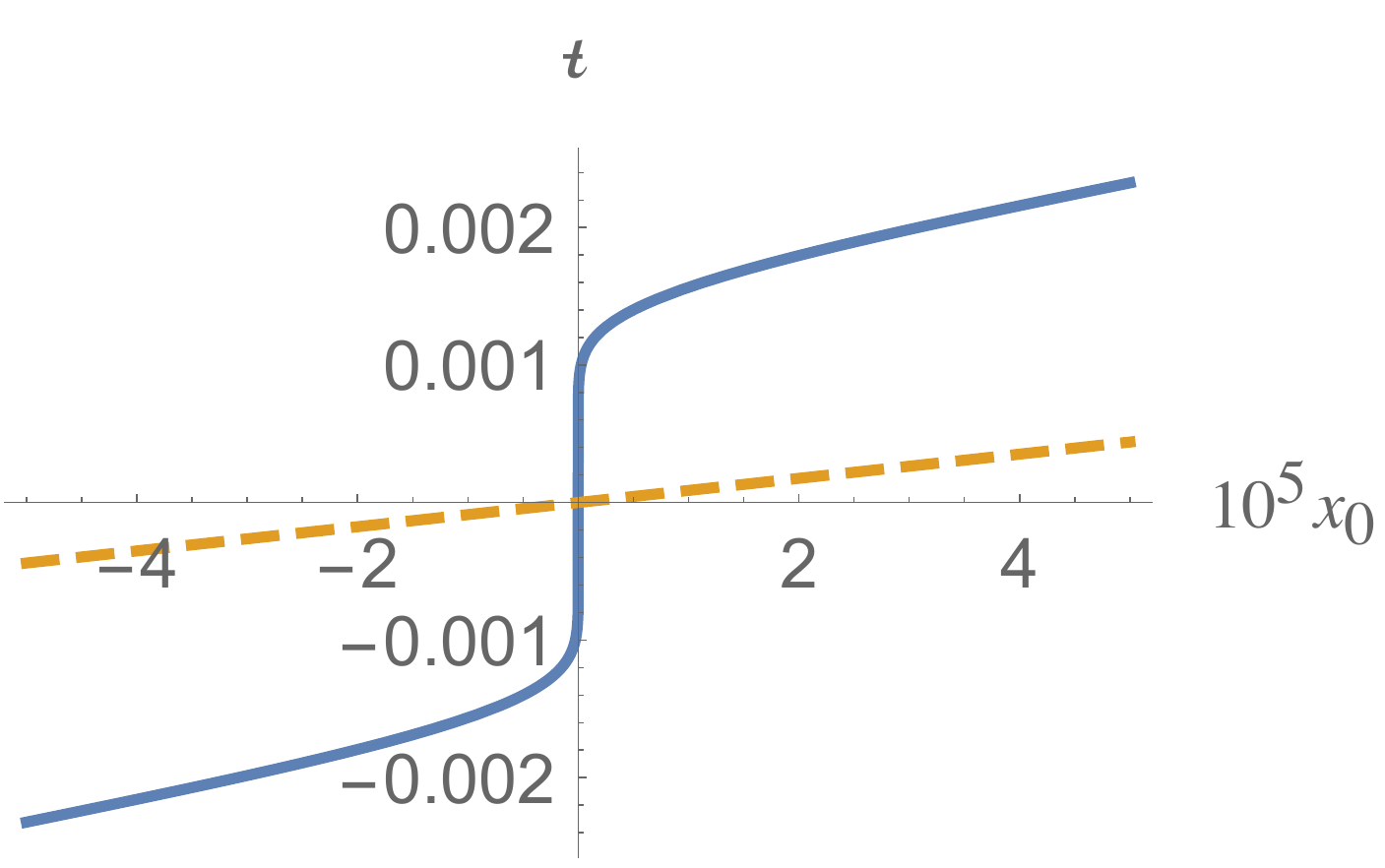} 
\caption{Bdy-\ref{itm:HKTbdy},  $\lambda = a = m =1$, taking $R = \abdy$: Rotational transform \emph{vs.} $x_0$ in the case $\mu_0 = 1.4/\abdy$, $\alpha = 0.005 > \alpha_{\rm thr}$ (see text). The dashed curve is for the unperturbed case $\alpha = 0$. }
\label{fig:transform}
\end{figure}
In Fig.~\ref{fig:transform} the rotational transform, \eqref{eq:transform}, is plotted, \emph{vs.} flux surface label. In this plot the amplitude parameter $\alpha = 0.005$, a value just above where current reversal ceases and the half-islands disappear (cf. Fig.~\ref{fig:sheetcurrent}). In this case the positive current in the sheet causes the rotational transform to change from the unperturbed resonant value $\iotabar = 0$ at $x = 0$,  jumping from a negative to a positive value across the current sheet. For smaller values of $\alpha$ the half-islands remove the discontinuity in $\iotabar$. This is because $|\grad\psi| = 0$ at the current reversal points, so, from \eqref{eq:qGS},  $q$ diverges as $x_0 \to 0$, locking $\iotabar$ to zero at $x_0 = 0$. However, the logarithmic nature of the singularity means this approach to zero manifests itself only at extremely small $x_0$, so the slope of $\iotabar(x_0)$ at the origin is so high that the plots for lower $\alpha$ look, to the eye, qualitatively the same as in Fig.~\ref{fig:transform} even at the very fine resolution in $x_0$ used in this figure.

\subsection{Sinusoidal rippled boundary condition}\label{sec:Bdy2}

\begin{figure}[htbp]
   \centering
		\includegraphics[width = 0.5\textwidth]{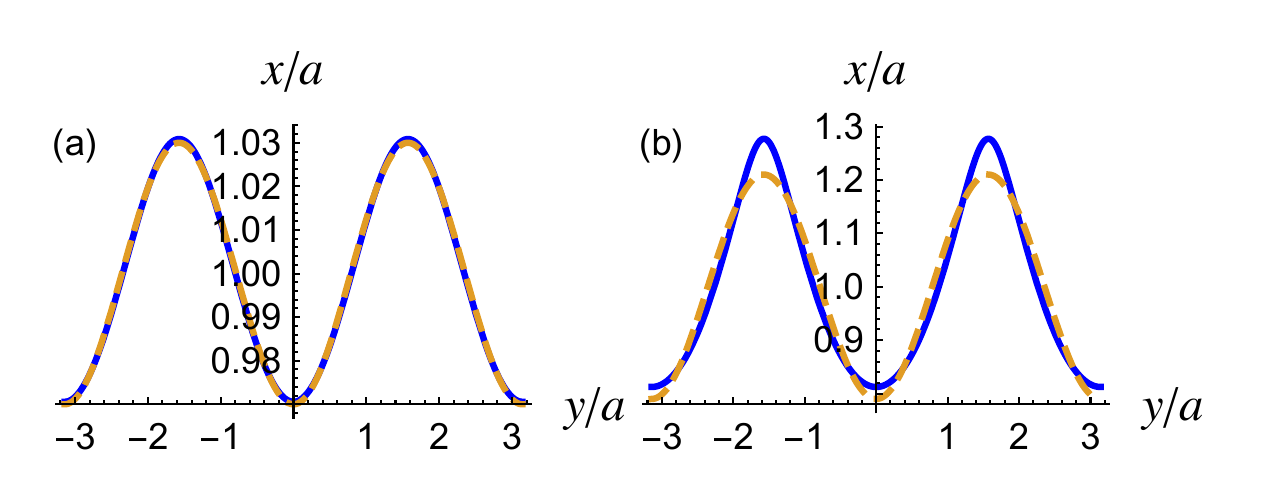} 
\caption{Panels (a) and (b) show boundaries generated by method Bdy-\ref{itm:HKTbdy} (blue online) for ripple amplitudes $\alpha = 0.03$ and $0.2$, respectively. For comparison, corresponding sinusoidal boundaries (dashed, orange online), as used in method Bdy-\ref{itm:sinbdy}, are also shown. In both panels, $m=2$, $\abdy\mu_0 = 0.2$. }
\label{fig:Bdy1Comparison}
\end{figure}

\begin{figure}[htbp]
   \centering
		\includegraphics[width = 0.5\textwidth]{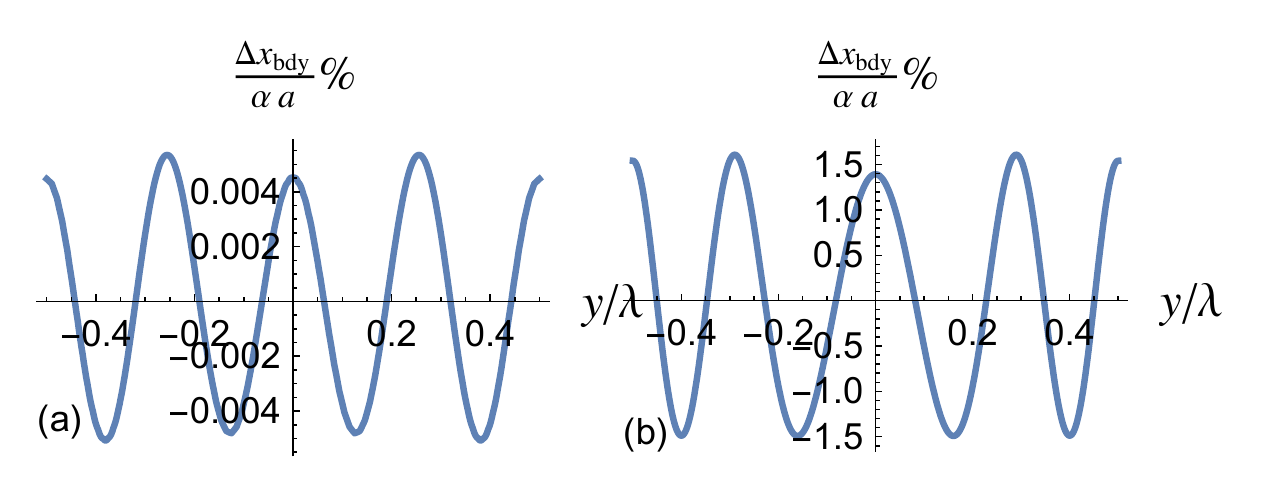} 
\caption{Percentage waveform errors using Bdy-\ref{itm:sinbdy} to fit the prescribed sinusoidal boundary in case $m=2$, $\abdy\mu_0 = 0.2$: (a) at small amplitude, $\alpha = 0.03$; (b) at larger amplitude, $\alpha = 0.21$.}
\label{fig:Bdy2errs}
\end{figure}

In Fig.~\ref{fig:Bdy1Comparison} we compare boundaries generated by method Bdy-\ref{itm:HKTbdy}, described in Sec.~\ref{sec:init}, with the corresponding sinusoidal boundaries defined by \eqref{eq:prescribedBdy}. For case (a), small amplitude ripple, method Bdy-1 produces a boundary indistinguishable from the target sinusoid, but for larger amplitude, case (b), strong second harmonic error is clear to the eye. 

In Fig.~\ref{fig:Bdy2errs} we plot the difference between the pure sinusoid defined by \eqref{eq:prescribedBdy} and boundaries generated by the Bdy-\ref{itm:sinbdy} method, (a) for small-amplitude ripple, $\alpha = 0.03$, and (b) for larger ripple, $\alpha = 0.21$. The $l$-sum in \eqref{eq:psigen}  was truncated after $l=3$ (hence the dominantly $l=4$ error) but even at $\alpha = 0.21$ the percentage waveform error of 1.5\% is tolerable for graphical work.

\begin{figure}[htbp] 
   \centering
		\includegraphics[width = 0.35\textwidth]{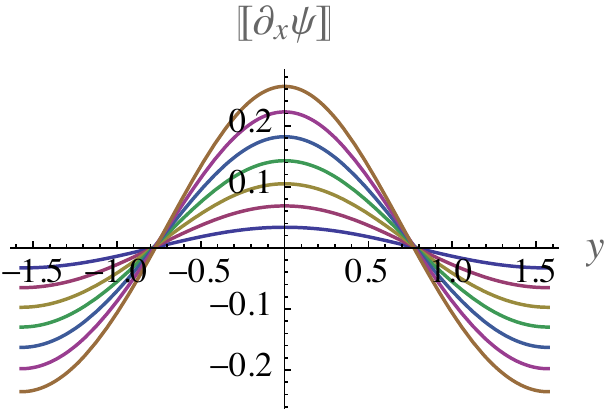} 
\caption{Bdy-\ref {itm:sinbdy},  $\lambda = 2\pi a/2$: Plots of the jump in the gradient of $\psi$, \eqref{eq:sheetstrength}, \emph{vs}. $y/\lambda$ for $m=2$, $\mu_0\abdy = 0.2$, and the set of amplitudes $\alpha$ given in the text, showing the occurrence of current-density reversal for all amplitudes in the set. }
\label{fig:Sinsheetcurrentm2}
\end{figure}

\begin{figure}[htbp] 
   \centering
		\includegraphics[width = 0.35\textwidth]{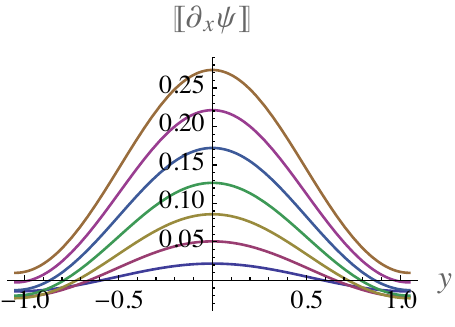} 
\caption{Bdy-\ref {itm:sinbdy},  $\lambda = 2\pi a/3$: Plots of the jump in the gradient of $\psi$, \eqref{eq:sheetstrength}, \emph{vs}. $y/\lambda$ for $m=3$, $\mu_0\abdy = 0.2$, and amplitudes given in the text, showing the occurrence of current-density reversal for all but the highest amplitude. }
\label{fig:Sinsheetcurrentm3}
\end{figure}

\begin{figure}[htbp] 
   \centering
		\includegraphics[width = 0.35\textwidth]{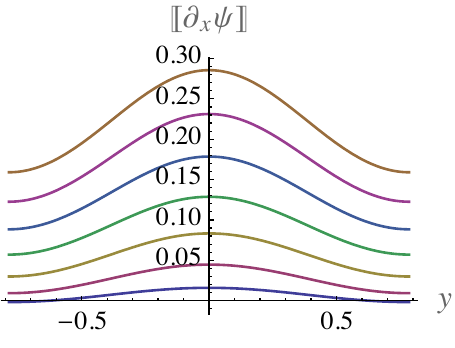} 
\caption{Bdy-\ref {itm:sinbdy},  $\lambda = 2\pi a/4$: Plots of the jump in the gradient of $\psi$, \eqref{eq:sheetstrength}, \emph{vs}. $y/\lambda$ for $m=4$, $\mu_0 \abdy = 0.2$ and amplitudes given in the text, showing the occurrence of current-density reversal only for the lowest amplitude. }
\label{fig:Sinsheetcurrentm4}
\end{figure}

Figures \ref{fig:Sinsheetcurrentm2}--\ref{fig:Sinsheetcurrentm4} plot the jump in the gradient of $\psi$, \eqref{eq:sheetstrength}, in the tokamak-relevant case $\mu_0\abdy = 0.2$, described after \eqref{eq:psiaconstraint}. The seven curves in each plot are for the set of amplitude values $\alpha = 0.03,0.06,0.09,0.12,0.15,0.18,0.21$, whose corresponding vertical-axis intersections run from bottom to top (color online). The three figures are for three values of ripple wave number $m$. The figures reveal the dramatic effect of $m$ on the phenomenon, illustrated in Figs.~\ref{fig:sheetcurrent}--\ref{fig:transform}, of the locking of rotational transform at the resonant value by half-island formation at small enough $\alpha$, transitioning beyond a threshold value of $\alpha$ to the  removal of the resonance  by the formation of a discontinuity in the rotational transform profile at the current sheet interface. 

The plots show there is qualitative transition in the rotational transform profile from strong resonance locking for $m\leq 3$ to no resonant locking, except at very small ripple amplitude, for $m \geq 4$. (This is consistent with the $m \approx 6$ results in Sec.~\ref{sec:Bdy1}, where resonance locking occurred only for extremely small $\alpha$.)

We interpret the stronger resonance locking at smaller $m$ as due to the greater penetration of the ripple perturbation from the boundary [$x = \xbdy(y)$] to the interface [$x = 0$] at longer wavelengths. This is a linear, $O(\alpha)$, effect and can easily be seen from the factor $1/\sinh \kappa_m\abdy \sim \exp(-\kappa_m\abdy)$ in \eqref{eq:sheetstrength} and the scaling $\gamma_{\rm S} \propto \alpha$, evident from Fig.~\ref{fig:gammaSshalpha}, showing $\gamma_{\rm S}$ can be ignored at linear order. On the other hand, it appears the $O(\alpha^2)$ d.c. response from the $\gamma_{\rm S}$ term is not so affected by the exponential decay of $\exp(-\kappa_m\abdy)$ and begins to dominate the linear response at larger $m$.

\section{Conclusion}\label{sec:conclude}
	In this paper we have used numerical calculations to give an exploratory overview of a geometrically simple application of dynamical MRxMHD in the adiabatic approximation, as well as expanding on the general formulation in \cite{Dewar_Yoshida_Bhattacharjee_Hudson_15} regarding the fundamental question of the appropriate definition for magnetic helicity in MRxMHD.
	
	Our calculations have confirmed the physical accessibility of the static, equilibrium solutions for RMP resonant states found by Loizu, Hudson \emph{et al.}, \cite{Loizu_etal_15b,Loizu_etal_16}, establishing the existence of a threshold RMP amplitude at which rotational-transform jumps across the resonantly excited current sheets occur. 
Having established this theoretical basis, the real test will be comparison with experiment, for instance revisiting the DIII-D reconstruction of RMP equilibria described in \cite[Sec.~IV. E ]{Hudson_etal_12b}, but using SPEC with appropriate entropy and magnetic helicity constraints. 
	
	While numerical calculations are a good way to explore the implications of a theory without being tied to particular parameter ranges, for a complete understanding they need to be complemented by analytical work in appropriate asymptotic regimes. In particular, the simple $\alpha$ scalings found empirically in this paper give confidence that an amplitude expansion could give an adequate understanding of RMP screening dynamics. Such an expansion procedure will be presented elsewhere.
	
	Another important area of research for which the HKT model is an ideal testbed for MRxMHD are investigations of reconnection mechanisms giving rise to the transition between perturbed states, fully shielded by the resonantly excited current sheet, to states with fully developed resonant islands. As this involves transfer of mass, entropy and magnetic flux between MRxMHD regions it is related to the field of helicity injection \cite{Jensen_Chu_84,Berger_Field_84,Finn_Antonsen_85}, whose study would require lifting the no-gaps restriction used in this paper. 

\section*{Appendices}
\appendix

\section{Magnetic helicity conservation with moving boundaries}
\label{sec:Kideal}

In this Appendix we establish that the gauge constraint on $\vrm{A}$ of conservation of surface loop integrals, mentioned in Sec.~\ref{sec:MRxMHD}, ensures time-independence of magnetic helicity in an ideal plasma region $\Omega^t$, with boundary $\partial\Omega^t$ dependent on time $t$. As the guiding principle in formulating MRxMHD is to invoke only constraints that are also appropriate in ideal MHD, this ideal constraint on $\vrm{A}$ is inherited as one of the foundational postulates of MRxMHD.

We need only the no-gaps tangential boundary condition \eqref{eq:Bn} and the ideal Ohm's Law 
\begin{equation}\label{eq:Ohm}
	\vrm{E} = -\vrm{v}\cross\vrm{B}\;,
\end{equation}
which, using Maxwell's equations, gives the equations of motion for $\vrm{B}$ and $\vrm{A},$
\begin{equation}\label{eq:Bdot}
	\frac{\partial\vrm{B}}{\partial t} = \curl(\vrm{v}\cross\vrm{B}) \;,
\end{equation}
and
\begin{equation}\label{eq:Adot}
	\frac{\partial\vrm{A}}{\partial t} = \vrm{v}\cross\vrm{B} - \grad\varphi \;,
\end{equation}
where $\vrm{A}$ is a single-valued vector field and $\varphi$ is a scalar potential, not assumed to be single-valued at this point. Dotting both sides of \eqref{eq:Adot} with $\vrm{v}$ and $\vrm{B}$ we find two differential equations for $\varphi$,
\begin{equation}\label{eq:Ohmdotv}
	\vrm{v}\dotv\grad\varphi = -\vrm{v}\dotv\frac{\partial\vrm{A}}{\partial t}\;,
\end{equation}
and
\begin{equation}\label{eq:OhmdotB}
	\vrm{B}\dotv\grad\varphi = -\vrm{B}\dotv\frac{\partial\vrm{A}}{\partial t}\;.
\end{equation}

Differentiating \eqref{eq:Helicity} and using the above assumptions we find [Supp] the time derivative of the magnetic helicity functional $K$ 
\begin{equation}\label{eq:Kdot}
\begin{split}
	2\upmu_0\frac{\d K}{\d t} &= \int_{\partial\Omega} \vrm{A}\dotv\vrm{B}\,\vrm{n}\dotv\vrm{v}\,\d S \\
			&\quad+ \int_{\Omega} \left[\frac{\partial\vrm{A}}{\partial t}\dotv\vrm{B}
			+ \vrm{A}\dotv\frac{\partial\vrm{B}}{\partial t} \right]\, \d V \\
			&=\sum_{l=1}^\nu \int_{S_l} \vrm{n}\dotv\vrm{B}\jump{\varphi}\,\d S \\
\end{split}
\end{equation}
It is thus seen that a \emph{sufficient} condition for invariance of $K$ is that $\jump{\varphi} \equiv 0$, i.e. that $\varphi$ be single valued everywhere.

We now test if single-valuedness is possible without contradicting Galilean invariance and, if so, what restrictions it places on the gauge of $\vrm{A}$. We assume the plasma is evolving under \eqref{eq:Bdot} with a prescribed velocity field $\vrm{v}$ from an initially integrable state with smoothly nested magnetic surfaces, which assumption will be hold for a finite time by the frozen-in flux argument \cite{Newcomb_58}. 

Consider first the Galilean invariance problem of a stationary state in the LAB frame as viewed from a moving frame, so the origin of the LAB frame appears to be moving with constant velocity $\vrm{v}_0$, hence $\vrm{v} = \vrm{v}_0 + \vrm{v}_{\rm L}$,  where subscripts L denote LAB frame fields viewed in the moving frame, and $\partial_t$ in LAB frame maps to $D_t \equiv \partial_t + \vrm{v}_0\dotv\grad$ in the frame of the observer. For example, \eqref{eq:Bdot} becomes
\begin{equation}\label{eq:BdotL}
	D_t\vrm{B}_{\rm L} = \curl(\vrm{v}_{\vrm L}\cross\vrm{B}_{\rm L}) \;,
\end{equation}
On the other hand, substituting $\vrm{v} = \vrm{v}_0 + \vrm{v}_{\rm L}$ in \eqref{eq:Bdot} we find
\begin{equation}\label{eq:BdotMov}
	D_t\vrm{B} = \curl(\vrm{v_{\rm L}}\cross\vrm{B}) \;.
\end{equation}
Comparing \eqref{eq:BdotL} and \eqref{eq:BdotMov} we see that $\vrm{B} = \vrm{B}_{\rm L}$, verifying Galilean invariance of $\vrm{B}$ in ``pre-Maxwell'' ideal MHD. (A Lorentz-invariant generalization of helicity has also recently been developed \cite{Yoshida_Kawazura_Yokoyama_14}.)

Assuming $D_t\vrm{A}_{\rm L} = 0$ and $\varphi_{\rm L}$ single valued, the LAB frame version of \eqref{eq:Adot} becomes
\begin{equation}\label{eq:AdotL} 
	\grad\varphi_{\rm L} = \vrm{v}_{\rm L} \cross\vrm{B}_{\rm L} \;,
\end{equation}
It is easily seen from the LAB version of \eqref{eq:OhmdotB} that $\varphi_{\rm L}$ must be constant on each magnetic surface, so, from the LAB version of \eqref{eq:Ohmdotv}, $\vrm{v}_{\rm L}$ is, like $\vrm{B}$, a tangential field on each magnetic surface.

Substituting $\vrm{v} = \vrm{v}_0 + \vrm{v}_{\rm L}$ in \eqref{eq:Adot} and using \eqref{eq:AdotL} we find
\begin{equation}\label{eq:AdotMov}
	\grad(\varphi - \varphi_{\rm L} - \vrm{v_0}\dotv\vrm{A}) = - D_t\vrm{A} \;.
\end{equation}
Taking line integrals of both sides on the magnetic surfaces around topologically distinct loops $C_l$, moving at the LAB frame velocity $\vrm{v}_0$ and cutting the corresponding surfaces of section $S_l$, we have
\begin{equation}\label{eq:AdotMovL}
	\oint_{C_l}\d\vrm{l}\dotv\grad(\varphi - \varphi_{\rm L} - \vrm{v_0}\dotv\vrm{A}) = - \oint_{C_l}\d\vrm{l}\dotv D_t\vrm{A} \;.
\end{equation}
Assuming single-valuedness of $\varphi$ (the other terms in the LHS integrand also being single-valued) the loop integrals on the LHS vanish. As the contours $C_l$ are stationary in the LAB frame we can commute $D_t$ outside the integral on the RHS to find
\begin{equation}\label{eq:GaliLoopInvar}
	\frac{\d}{\d t}\oint_{C_l}\d\vrm{l}\dotv \vrm{A} = 0\;.
\end{equation}
Thus the loop integrals of $\vrm{A}$ on magnetic surfaces are time-invariant in all frames, which is consistent with Galilean invariance of $\vrm{A}$: $\curl\vrm{A} = \curl\vrm{A}_{\rm L}$ is solved by $\vrm{A} = \vrm{A}_{\rm L} + \grad\chi$, where $\chi$ is an arbitrary but \emph{single-valued} gauge potential. This confirms that single-valuedness of $\varphi$ is consistent with Galilean invariance of $K$ for systems that are stationary in the LAB frame. 

We now consider systems that are not stationary in any frame, i.e. $\vrm{v}$ is an arbitrary function of time and space, advecting the magnetic surfaces. Can we show that single-valuedness of $\varphi$ always implies time-invariance of $\oint_{C_l}\d\vrm{l}\dotv \vrm{A}$ around magnetic surfaces (particularly the plasma boundary) even for loops not enclosing the plasma? If so, this is the boundary condition consistent with conservation of $K$.

We begin with the advective
 \footnote{We could split $\vrm{v}$ into flows normal and tangential to the magnetic surfaces to prevent the $C_l$ 
  being advected around the magnetic surfaces, but for simplicity we do not do this as it is easy to see all such loop 
  integrals on a given flux surface are equal. Thus tangential advection of the $C_l$ does not affect invariance.}
form of \eqref{eq:Adot} [cf. e.g. eq.~(1b) of \cite{Dewar_70}].
\begin{equation}\label{eq:AdotAdv}
\begin{split}
	\frac{\d\vrm{A}}{\d t} &= (\grad\vrm{A})\dotv\vrm{v} - \grad\varphi \\
	&= - (\grad\vrm{v})\dotv\vrm{A} - \grad(\varphi - \vrm{v}\dotv\vrm{A})  \;,
\end{split}
\end{equation}
where $\d/\d t \equiv \partial_t + \vrm{v}\dotv\grad$. We also need the advection of a line element $\d\vrm{l} = \vrm{r}(\vrm{r}_0 + \d\vrm{l}_0,t) - \vrm{r}(\vrm{r}_0,t) = \d\vrm{l}_0\dotv\grad_0\vrm{r}(\vrm{r}_0,t)$, where $\d\vrm{l}_0$ is an infinitesimal displacement in the initial position $\vrm{r}_0$ of a fluid element some time before the present
\begin{equation}\label{eq:dlAdv}
\begin{split}
	\frac{\d}{\d t}\d\vrm{l} &= \vrm{v}(\vrm{r}(\vrm{r}_0 + \d\vrm{l}_0,t)) - \vrm{v}(\vrm{r}(\vrm{r}_0,t)) \\
	&=  \d\vrm{l}_0\dotv\grad_0\vrm{r}(\vrm{r}_0,t)\grad\vrm{v}(\vrm{r},t)) \\
	&= \d\vrm{l}\dotv\grad\vrm{v}  \;,
\end{split}
\end{equation}

Then
\begin{equation}\label{eq:LoopInvariance}
\begin{split}
	\frac{\d}{\d t}\oint_{C_l}\d\vrm{l} \dotv \vrm{A} 
	&= \oint_{C_l}\left[\left(\frac{\d}{\d t}\d\vrm{l}\right)\dotv \vrm{A} 
	+ \d\vrm{l} \dotv \frac{\d\vrm{A}}{\d t}\right] \\
	&= \oint_{C_l}\left\{\d\vrm{l}\dotv(\grad\vrm{v}) \dotv \vrm{A} \right.\\
	&\left.\quad - \d\vrm{l} \dotv \left[(\grad\vrm{v})\dotv\vrm{A} + \grad(\varphi - \vrm{v}\dotv\vrm{A})\right]\right\} \\
	&= 0 
\end{split}
\end{equation}
if and only if $\varphi$ is single valued, which is also the condition for $K$ to be time-invariant, so the \emph{full} helicity constraint condition in $\Omega_i$ is conservation of $K_i$ \emph{and} constancy of $\oint\d\vrm{l}\dotv\vrm{A}$ around all topologically distinct loops on each disjoint component of the boundary $\partial\Omega_i$.

Note an \emph{additional} loop integral constraint:  If $\Omega_i$ and $\Omega_j$ are neighboring regions, the corresponding loop integrals  $\oint_{\pm}\d\vrm{l}\dotv\vrm{A}$ on the two sides $\pm$ of the common boundary $\Omega_{i,j}$ are constrained to be \emph{equal} because finiteness of $\vrm{B}$ requires there be vanishing magnetic flux trapped within the common interface.

\section{Vacuum Helicity}
\label{sec:vacHel}

In this Appendix we illustrate the fact that vacuum helicity is not geometrically invariant by showing it is not invariant even in slab geometry (if poloidal vacuum field is included). This shows that the Finn--Antonsen \cite{Finn_Antonsen_85} form of the magnetic helicity (equivalent to the Jensen--Chu \cite{Jensen_Chu_84} relative helicity when there are no gaps in the perfectly conducting boundaries) is not appropriate in MRxMHD.

Following \cite{Yoshida_Dewar_12} we define the harmonic (vacuum) component $\BH$ of a Beltrami field $\vrm{B}$ in an annular toroid as the curl-free ($\mu = 0$) component carrying the toroidal and poloidal fluxes. 
Specifically, consider $\Omega_{+}$, for which the toroidal ($z$-directed) flux is $2\pi\abdy^2\Fav$ [$F$ being constant when $\mu = 0$, from \eqref{eq:BeltramiGS2}] and the poloidal ($y$-directed) flux is, from \eqref{eq:polflux}, $2\pi \psibdy R$. Then
\begin{equation}\label{eq:GSBrepH}
	\BH = \Fav\,\esub{z}  + \esub{z} \cross \grad \psi^{\rm H}
\end{equation}
where the general form of $\psi_{\rm H}(x,y)$ such that $\nabla^2\psi = 0$ and $\psi_{\rm H}(0,y) = 0$ is [cf. \eqref{eq:psigen} with $\mu = 0$],
\begin{equation}\label{eq:psigenH}
\begin{split}
	\psi^{\rm H}(x,y) &=  d^{\rm H}_0\,|x| 
	 + \sum_{l=1}^{\infty} d^{\rm H}_{l m} \cos \frac{l m y}{\abdy}\, \sinh \left|\frac{l m x}{\abdy}\right| \;.
\end{split}
\end{equation}
with the corresponding vector potential [cf. \eqref{eq:GSArx}]
\begin{equation}\label{eq:AH}
	\AH =   -\psi^{\rm H}\esub{z} +\esub{z}\cross\grad \left(\half \Fav x^2\right) \;.
\end{equation}
The coefficients $d^{\rm H}$ are to be chosen so that the Dirichlet boundary condition
\begin{equation}\label{sec:psiHbdy}
	\psi^{\rm H}(\xbdy(y),y) = \psibdy \quad \forall\,y
\end{equation}
is satisfied, in order to conserve poloidal flux.

Then we define the \emph{vacuum helicity} \cite{Jensen_Chu_84,Berger_Field_84,Finn_Antonsen_85} analogously to $K_i$, \eqref{eq:Helicity}, as [Supp] 
\begin{equation}\label{eq:HelicityHdef}
	\begin{split}
		K^{\rm H}_{+} &\equiv \int_{\Omega_{+}} \frac{\AH\dotv\BH}{2\muSI} \, \d V \\
		&= \int_{\Omega_{+}} \frac{
			\left[ -\Fav\psi^{\rm H} 
			+ \left(\grad\half \Fav x^2\right)\dotv \grad \psi^{\rm H}) \right]
			}{2\muSI} \, \d V \;.
	\end{split}
\end{equation}
Integration by parts then gives [Supp] 
\begin{equation}\label{eq:HelicityH}
	\begin{split}
	K^{\rm H}_{+} 
		&= \int_{\Omega_{+}} \frac{
			\left[ -\Fav\psi^{\rm H} 
			+ \Fav \partial_x(x\psi^{\rm H}) \right]
			}{2\muSI} \, \d V \\
		&= \frac{\Fav}{2\muSI}\left[
			 -2\int_{\Omega_{+}}\psi^{\rm H} \, \d V
			+ 2\pi\abdy^2\psibdy
			\right] \;,
	\end{split}
\end{equation}
using the area constraint \eqref{eq:meanxbdy}.

The term $2\pi\abdy^2\psibdy$ is invariant under changes in $\xbdy(y)$. However, there appears no reason for the integral $\int_{\Omega_{+}}\!\!\psi^{\rm H} \, \d V$, evaluated from \eqref{eq:psigenH} as [Supp] 
\begin{equation}\label{eq:psiHint}
	\begin{split}
		\int_{\Omega_{+}}\psi^{\rm H} \, \d V
	 &= \int_{-\pi\abdy}^{\pi\abdy}\d y  \left[
		\frac{1}{2} d^{\rm H}_0 \xbdy^2(y) \right. \\
	  \mbox{}+\sum_{l=1}^{\infty} &
	 \left.  d^{\rm H}_{l m} \cos \left(\frac{l m y}{\abdy}\right)
	 \left(\cosh \left|\frac{l m \xbdy(y)}{\abdy}\right| - 1\right)
	 \right] \;,
	\end{split}
\end{equation}
to be invariant in general. As the vacuum helicity \eqref{eq:HelicityH} includes this integral we conclude that  $K^{\rm H}_{+}$ is not in general invariant and therefore not suitable for defining a relative helicity that is conserved under deformations in boundary shape.

\section{Scalar Variational Principle}\label{sec:psiBeltrami} 

For a variational approach to deriving the Grad--Shafranov form of the Beltrami equation see the online Supplement [Supp]. 

\section*{Supplementary material}

See online supplementary material [Supp] (provided as an ancillary file in this arXiv version) for further details relevant to this paper: more detailed derivations and Appendix~\ref{sec:psiBeltrami}. 

\section*{Acknowledgments}
One of the authors (RLD) gratefully acknowledges the support of Princeton Plasma Physics Laboratory and The University of Tokyo during development of the concepts in this paper through collaboration visits over a number of years, and some travel support from Australian Research Council grant DP11010288. The work of SRH and AB was supported under US DOE grant DE-AC02-09CH11466 and that of ZY was supported under JSPS grant KAKENHI 23224014. The numerical calculations and plots were made using Mathematica 10, \cite{Mathematica10}.

\bibliography{RLDBibDeskPapers}

\end{document}